\documentclass[12pt,a4paper,final]{article}
\usepackage[utf8]{inputenc}
\usepackage{amsmath,amsthm,amsfonts,amssymb}
\usepackage{makeidx}
\usepackage{graphicx}
\usepackage{tikz}
\usetikzlibrary{calc,positioning,shapes.geometric,shapes.symbols,shapes.misc}
\usepackage[linesnumbered,algoruled,boxed,lined]{algorithm2e}
\usepackage{hyperref}

\newtheorem{thm}{Theorem}
\newtheorem{lem}{Lemma}

\newtheorem{Open problem}{Open Problem}

\newtheorem{prop}{Proposition}

\newtheorem{defn}{Definition}
\newtheorem{prob}{Problem}
\newtheorem{conj}{Conjecture}
\newtheorem{ex}{Example}

\newcommand{\F}{\mathbb{F}}
\newcommand{\bs}{\mathbf{s}}
\newcommand{\bv}{\mathbf{v}}
\newcommand{\calF}{\mathcal{F}}
\makeatletter
\newcommand{\tpmod}[1]{{\@displayfalse\pmod{#1}}}
\makeatother
\begin{document}
	\title{A Survey on Complexity Measures for Pseudo-Random Sequences}
	\author{Chunlei Li \\
		\small	Email: chunlei.li@uib.no \\
			\small Department of Informatics, 
			\small University of Bergen, 5020 Bergen, Norway		
	}
	\date{}
	
	\maketitle
	
	\begin{abstract}
		Since the introduction of the Kolmogorov complexity of binary sequences in the 1960s, 
		there have been significant advancements in the topic of complexity measures for randomness assessment, which are of fundamental importance in
		theoretical computer science and of practical interest in cryptography. This survey reviews notable
		research from the past four decades on the linear, quadratic and maximum-order complexities of pseudo-random sequences
		and their relations with Lempel-Ziv complexity, expansion complexity, 2-adic complexity, and correlation measures.		
	\end{abstract}


	\section{Introduction}
	
	{ Cryptography and security applications make extensive use of random bits, such as keys and initialization vectors in encryption and nonces in security protocols.
		The generation of random bits should be designed with the security goal of \textit{indistinguishability} from an ideal randomness source, which generates identically distributed and independent uniform random bits with full entropy (i.e., one bit of entropy per bit).
		However, this security goal is challenging to achieve. The list of real-world failures, where a random bit generator (RBG) is broken and the security of the reliant application crumbles with it, continues to grow \cite{ristenpart2010, Woodage2019, DualEC, Hannah2024}.
		There are two fundamentally different strategies for designing RBGs. One strategy is to produce bits non-deterministically, where every bit of output is based on a physical process that is unpredictable; the other strategy is to compute bits deterministically using an algorithm from an initial value that contains sufficient entropy to provide an assurance of randomness. This class of RBGs are known as pseudo-random bit generators (PRBGs) or deterministic random bit generators (DRBGs) \cite{Barker2015Jun}. 
		Due to their deterministic nature, PRBGs produce sequences of pseudo-random (rather than random) bits. 
		Real-world PRBGs usually employ cryptographic primitives, such as stream ciphers, block ciphers, hash functions  and elliptic curves, as their basic building blocks. 
		For instance, the revised NIST SP 800-90A standard \cite{Barker2015Jun} recommends Hash-DRBG, HMAC-DRBG and CTR-DRBG based on approved hash functions and block ciphers.
		It is worth mentioning that, as pointed in \cite{Woodage2019}, the NIST SP 800-90A standard is not free from controversy: the disgraced algorithm DualEC-DRBG in the standard was reported to contain a back door \cite{DualEC};
		the recommended DRBGs, when supporting a variety of optional inputs and parameters,  do not fit cleanly into the usual security models of PRBGs; there is no formal competition of the standarisation process, and there is a limited amount of formal analysis of those recommended DRBGs.
		Although not included in the NIST standard, PRBGs that use stream ciphers based on feedback shift registers (FSRs) as building blocks have several advantages,
		including simple structure of operations of addition and multiplication, fast and easy hardware implementations in almost all computing devices,  good statistic characteristics (long period, balance, run properties, etc.) in output sequences \cite{Sarkar2006, Kuznetsov2022}.
		In addition, the output sequence from other PRBGs will ultimately become periodic and therefore can be produced by certain FSRs. 
		This provides another perspective for investigating the security strength of output sequences from PRBGs.
	} 
	
	{ Consider a pseudo-random sequence $\bs=s_0s_1s_2\dots$ generated from a PRBG.
		A natural question arises: how should one evaluate the \textit{randomness} of the sequence $\bs$?}
	Researches on this problem can be traced back to the early 20th century. As early as 1919 Mises \cite{Mises1919} initiated the notion of random sequences based on his frequency theory of 
	probability. Later Church \cite{Church1940} pointed out the vagueness of the second condition on randomness in \cite{Mises1919} and proposed
	a less restricted but more precise definition of random sequences: 
	an infinite binary sequence $\bs$ is a random sequence if it satisfies the following two conditions: 
	{1) assume $f(r,\bs)$ denotes the number of 1's among the first $r$ terms of $\bs$, then the sequence $f(r,\bs) / r$ for $\bs$ approaches a limit $p$ as $r$ approaches infinity;
		2) for a sequence $\mathbf{b}$ with $b_1=1, b_{n+1}=2 b_n+s_n$ and any effectively calculable function $\phi$,
		if the integers $n$ such that $\phi(b_n)=1$ form an infinite sequence $n_1, n_2, \cdots$,  then the sequence $f(r,\mathbf{c}) / r$ for $\mathbf{c}=s_{n_1}s_{n_2}s_{n_3}\dots$ approaches the same limit $p$ as $r$ approaches infinity.}
	Despite the emphasis of calculability of $\phi$ in Condition 2), one can see that it is hardly (if not at all) possible to 
	test the randomness of a sequence by this definition. It could be used in randomness test by negative outcomes, namely, 
	a binary sequence $\bs$ is not random if one can find certain effectively calculable function $\phi$, for which Condition 2) cannot hold.
	In order to justify a proposed definition of randomness, one has to show that the sequences, which are random in the stated sense, 
	possess various properties of stochasticity with which we are acquainted in probability theory.
	Inspired by the properties obeyed by the maximum-length sequences, Golomb proposed the \textit{randomness postulates} of binary sequences \cite{Golomb2017}: balancedness, run property\footnote{A subsequence of consecutive 1's/0's in a sequence is termed a run. A maximum-length sequence of length $2^m-1$
		contains one run of $m$ 1's, one run of $(m-1)$ 0's and $2^{k}$ runs of $m-2-k$ 1's and 0's for $k=0,1,\dots, m-3$.}  and ideal autocorrelation. 
	{ Kolmogorov \cite{Kolmogorov1998Nov} proposed the notion of \textit{complexity} to test the randomness of a sequence $\bs$, 
		which is defined as $\min_{A(\mathbf{x})=\bs}{\rm len}(\mathbf{x})$, the length of the shortest $\mathbf{x}$ that can produce $\bs$ by a universal Turing machine program $A$.
		This notion is later known as the Kolmogorov complexity in the literature.
		Along this line further developments were made in \cite{MartinLoef1966, Schnorr1973} and summarized in Knuth's famous book series  the Art of Computer Programming \cite{Knuth1997}. }
	
	Rather than considering an abstract Turing machine program that generates a given sequence, the model of using FSRs to generate a given sequence had attracted considerable attentions. This model has two major advantages: firstly, all sequences that are generated by a finite-state machine in a deterministic manner are ultimately periodic, and as such, can be produced by certain finite-state shift registers; secondly, it is comparatively easier and more efficient to identify the shortest FSRs 
	producing a given sequence, either of a finite length or of infinite length with a certain period. 
	Due to its tractability, \textit{linear complexity} of a specific sequence $\bs$, which uses linear FSRs (LFSRs) as the algorithm $A$ in the Kolmogorov complexity, 
	is particularly appealing and has been intensively studied. The linear complexity of a given sequence $\bs$ can be efficiently calculated by the Berlekamp-Massey algorithm \cite{Berlekamp1968,Massey1969},
	and the stochastic behavior of linear complexity for finite-length sequences and periodic sequences can be considered to be fairly well understood \cite{Rueppel1986,Niederreiter1988,Niederreiter2003}.
	The good understanding of linear complexity of sequences is an important factor for its adoption in the NIST randomness test suite \cite{Bassham2010}.
	Note that extremely long sequences with large linear complexity can usually be generated by much short FSRs with certain nonlinear feedback function.
	Consequently, more figures of merit to judge the randomness of sequences, such as maximum-order complexity \cite{Jansen1989}, 2-adic complexity \cite{Klapper1997}, quadratic complexity \cite{Chan1990}, expansion complexity \cite{Diem2012} and their variants, have been also explored in the literature.
	
	This paper will survey the development of complexity measures used to assess the randomness of sequences. 
	It is important to note that this paper does not intend to offer a comprehensive survey of this broad topic. 
	Instead it aims to provide a preliminary overview of the topic with technical discussions to certain extent, focusing on mainly complexities within the domain of FSR that align with the author's research interests.
	To this end, I delved into significant findings on the study of FSR-based complexity measures, particularly on the algorithmic and algebraic methods in computation, statistical behavior, theoretical constructions, relations of those complexity measures. Research papers on this topic presented at flagship conferences in cryptography, such as ASIACRYPT, CRYPTO and EUROCRYPT, published at prestigious journals with IEEE, Springer, Science Direct, etc, as well as some well-recognized books since 1970s constitute the major content of this survey. Readers may refer to
	relevant surveys on this topic with different focuses by Meidl and Niederreiter \cite{Meidl2002a, Niederreiter2003}, Topuzo\v{g}lus and Winterhof \cite{Topuzoglu}. 
	
	The remainder of this paper is organized as follows: Section \ref{sec_preliminary}  recalls the 
	basics for feedback shift registers, which lays the foundation for the discussions in subsequent sections. 
	Section \ref{sec_lc} reviews important developments of the theory of linear complexity profile for random sequences, Section \ref{sec_qc} discusses
	a mathematical tool used to efficiently calculate the quadratic complexity of sequences and
	Section \ref{sec_nc}  surveys computational methods, the statistical behavior for maximum-order complexity, and theoretical constructions  
	of periodic sequences with largest maximum-order complexity. In Section \ref{sec_relation} known relations among complexity measures 
	of sequences are briefly summarized. Finally, conclusive remarks and discussion are given in Section \ref{sec_conclusion}.

	\section{Feedback Shift Registers} \label{sec_preliminary}

	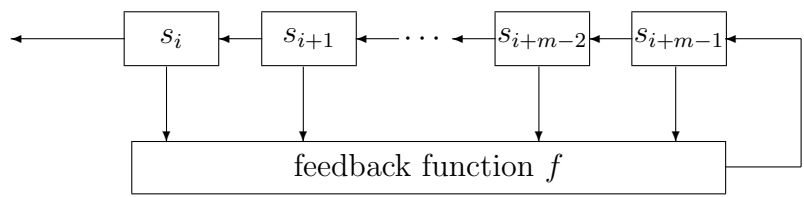
\begin{figure}[!ht]
		\begin{center}
			\setlength{\unitlength}{1mm}
			\begin{picture}(110,35)
			\put(15,25){\framebox(12.3,7)[c]{$s_i$}}\put(33.1,25){\framebox(12.3,7)[c]{$s_{i+1}$}}
			\put(63.8,25){\framebox(12.3,7)[c]{$s_{i+m-2}$}}\put(81.8,25){\framebox(12.3,7)[c]{$s_{i+m-1}$}}
			\put(51,25){\makebox(7,7)[c]{$\cdots$}}
			\put(33.1,28.5){\vector(-1,0){5.7}}\put(51.1,28.5){\vector(-1,0){5.7}}
			\put(63.8,28.5){\vector(-1,0){5.7}}\put(81.8,28.5){\vector(-1,0){5.7}}
			\put(15,28.5){\vector(-1,0){15}}
			\put(16,8){\framebox(78,6.8)[c]{feedback function $f$}} 
			\put(20.5,24.8){\vector(0,-1){10}}\put(38.5,24.8){\vector(0,-1){10}}
			\put(69.5,24.8){\vector(0,-1){10}}\put(87.5,24.8){\vector(0,-1){10}}
			\put(94,11.5){\line(1,0){10}}\put(104,11.5){\line(0,1){17}}\put(104,28.5){\vector(-1,0){10}}
			\end{picture}
		\end{center}
		\vspace{-1.2cm}\caption{An $m$-stage FSR with feedback function $f$}
		\label{Fig1}
	\end{figure}
	
	Let $\F_q$ be the finite field of $q$ elements, where $q$ is an arbitrary prime power. 
	For a positive integer $m$, an $m$-stage \textit{feedback shift register} (FSR) is a clock-controlled circuit consisting of $m$ consecutive storage units and a feedback function $f$ as displayed in Figure \ref{Fig1}.
	Starting with an initial state $s_{0}s_{1}\dots s_{m-1}$ over $\F_q$, the states in the FSR will be updated by a clock-controlled transformation as follows:
	\begin{equation}\label{trans_A}
	\calF: s_{i}s_{i+1}\dots s_{i+m-1} \longmapsto  s_{i+1}\dots s_{i+m-1}s_{i+m}, \, i \geq 0,
	\end{equation} where $$s_{i+m}=f(s_{i},s_{i+1},\dots,s_{i+m-1}),$$ and the leftmost symbol for each state will be output. In this way an FSR produces a sequence $\bs=s_0s_{1}s_2\dots$ based on each initial state $s_{0}s_{1}\dots s_{m-1}$ and its  feedback function $f$. 
	{ The output sequence from an FSR, known as an FSR sequence, can be equivalently expressed as a sequence of states $\underline{\bs}_i=s_is_{i+1}\dots s_{i+m-1}$,  with the relation 
		$\underline{\bs}_i = \calF(\underline{\bs}_{i-1})= \cdots = \calF^i(\underline{\bs}_0)$ for $i\geq 0$.
		When $\underline{\bs}_p =\calF^p(\underline{\bs}_0)= \underline{\bs}_0$ for the least integer $p\geq 1$, we obtain a cycle of states $\underline{\bs}_0,\dots \underline{\bs}_{p-1}$, or equivalently a sequence $s_0\dots s_{p-1}\dots$ of least period $p$. 
		In his influential book \cite{Golomb2017}, Golomb intensively studied
		the properties of FSR sequences from both linear feedback functions and nonlinear feedback functions. 
		Readers can refer to \cite{Golomb2017} for a comprehensive understanding of FSR sequences.
		%
	}
	
	\smallskip
	
	For an $m$-stage FSR with a feedback function $f$ and an initial state $s_0s_1\dots s_{m-1}$, if we know
	$n$ consecutive $s_is_{i+1}\dots s_{i+n-1}$ in the output $\bs$, then we obtain $n-m$ equations 
	\begin{equation}\label{eq_FSR_recurrence}
	\begin{array}{rcl}
	f(s_{i}, \dots, s_{i+m-1}) &=&s_{i+m}  \\
	f(s_{i+1}, \dots, s_{i+m}) &=&s_{i+m+1}  \\
	& \vdots&
	\\
	f(s_{i+n-m-1}, \dots, s_{i+n-2}) &=&s_{i+n-1}.  \\
	\end{array}
	\end{equation}
	When the feedback function $f$ is a linear, namely, $f(x_0,\dots, x_{m-1})=a_{1}x_{m-1} + a_2x_{m-2} + \cdots + a_mx_0 $, and $n\geq 2m$, one can uniquely determine its $m$ unknown coefficients $a_i$ from the above $n-m$ linear equations. 
	When the feedback function $f$ is nonlinear, the number of terms in $f$ increases significantly. For instance, for $q=2$ a binary feedback function $f$ of degree $r$ has $\sum_{d=0}^{r}\binom{n}{d}$ possible terms. 
	The above equations in Eq. \eqref{eq_FSR_recurrence} become more difficult to analyze. The equations are not necessarily linearly independent and a lot more variables are involved.
	Consequently, more observations and techniques are required in the analysis of these equations.
	In the following we will review some results on FSR sequences when the feedback function is linear, quadratic and arbitrarily nonlinear, and their relations with other complexity measures.

	We will consider both finite-length sequences and infinite-length sequences with certain finite period. Throughout what follows, we will use $\bs$ to denote a generic sequence with terms from certain alphabet, $\bs_n=s_0s_1\dots s_{n-1}$ to denote a sequence with 
	length $n$ that is not a repetition of any shorter sequence, $\bs_n^k$ to denote a sequence of $k$ repetition of $\bs_n$, and $\bs_n^{\infty}$ to denote infinite repetitions of $\bs_n$, indicating that $\bs_n^{\infty}$ has least period $n$.
	
	

	\section{Linear Complexity}\label{sec_lc}
	Given an FSR, when its feedback function $f$ is a linear function on the input state, namely, it is associated with the following linear recurrence: 
	\begin{equation}\label{eq:linear_recurrence}
	s_{i+m}+a_1s_{i+m-1}+\cdots+a_{m-1} s_{i+1}+a_m s_i = 0 \text { for } i=1,2, \ldots,
	\end{equation} where $a_1,\dots, a_{m}$ are taken from $\F_q$, 
	it is termed linear FSR and the output sequence $\bs$ is called an LFSR sequence of order $m$. 
	The polynomial $$f(x) = x^m + a_{m-1}x^{m-1} + \dots + a_{1}x + a_0,$$ with $a_0=1$, is 
	called the \textit{feedback or characteristic polynomial} of $\bs$.
	The zero sequence $00\dots$ is viewed as an LFSR sequence of order 0. 
	
	\begin{defn}
		Let $\bs$ be an arbitrary sequence of elements of $\F_q$ and let $n$ be a non-negative integer. Then the \textit{linear complexity} $L_n(\bs)$ is defined as the 
		length of the shortest LFSR that can generate  $\bs_n=s_0s_1\dots s_{n-1}$.	The sequence $L_0(\bs), L_1(\bs), L_2(\bs), \dots, $ is called the \textit{linear complexity profile} of the sequence $\bs$.
	\end{defn}
	
	It is clear that $L_n(\bs)\leq L_{n+1}(\bs)$ and $0\leq L_n(\bs)\leq n$ for any integer $n$ and sequence $\bs$. Hence the linear complexity profile of $\bs$ is a nondecreasing sequence of nonnegative integers. 
	Two extreme cases for $L_n(\bs)$ correspond to highly non-random sequences $\bs$ whose first $n$ elements $s_0s_1\dots s_{n-1}$ are either $\overbrace{0\dots0}^n$ or $\overbrace{0\dots 0}^{n-1} s_{n-1}$ with $s_{n-1}\neq 0$,
	which has $L_n(\bs) = 0$ or $L_n(\bs)=n$, respectively.
	
	The linear complexity of a sequence $\bs$ can be efficiently calculated by the well-known Berlekamp-Massey algorithm \cite{Berlekamp1968,Massey1969}, which also returns the linear feedback function generating $\bs$.
	In the Berlekamp-Massey algorithm, at each step $n-1$ with the current linear recurrence for $s_0s_1\dots s_{n-2}$, a discrepancy is calculated to assess whether the 
	linear recurrence holds for the extended sequence $s_0s_1\dots s_{n-2}s_{n-1}$: 
	when the discrepancy is zero, indicating that the current linear recurrence indeed holds for $s_0s_1\dots s_{n-2}s_{n-1}$, then move on to $n$; 
	when the discrepancy is nonzero, indicating that a new (and possibly longer) linear recurrence is needed, the linear recurrence is then updated and the linear complexity is updated as 
	\begin{equation}\label{eq_lc_update}
	L_{n}(\bs)=\max\{L_{n-1}(\bs), n-L_{n-1}(\bs)\}.
	\end{equation}
	{ Gustavson \cite{Gustavson1976May} showed that the above process requires $O(n^2)$ multiplication/addition operations for calculating $L_n(\bs)$.
		Dornstetter in \cite{Dornstetter1987May} pointed out the equivalence between this procedure of calculation and the Euclidean algorithm. 
	}
	
	Rueppel \cite{Rueppel1986} investigated the varying behavior of the discrepancy $\Delta_t$, $1\leq t <n$, for a binary sequence of length $n$, and characterized the linear complexity profile of random sequences in the following propositions.
	\begin{prop}
		The number of sequences of length $n$ and linear complexity $l$, denoted by $N_n(l)$, satisfies the following recursive relation
		\[
		N_n(l)= \begin{cases}
		N_{n-1}(l), &  0 \leq l <\frac{n}{2},
		\\ 2 N_{n-1}(l), & l=\frac{n}{2}, \\
		2 N_{n-1}(l)+N_{n-1}(n-l), & \frac{n}{2}<l<n, 
		\end{cases}
		\] and it, starting from $N_1(0) = N_1(1) =1$, can be given by 
		$$
		N_n(l)= \begin{cases}2^{\min \{2 n-2 l, 2 l-1\}}, &0<l\leq n, \\ 1, & 0=l<n.\end{cases}
		$$
	\end{prop}
	
	\begin{prop} The expected linear complexity  $L_n$ of  binary random sequences  of length $n$ drawn from a uniform distribution is given by 
		$$
		E\left[L_n\right]=\frac{n}{2}+\frac{4+R_2(n)}{18}-2^{-n}\left(\frac{n}{3}+\frac{2}{9}\right),
		$$
		and the variance of the linear complexity is given by 
		$$
		\begin{aligned}
		\operatorname{Var}\left[L_n\right] & =\frac{86}{81}-2^{-n} \frac{3n(14-R_2(n))-(82-2R_2(n))}{81}
		-2^{-2 n}\frac{9n^2 + 12n +4}{81},	 
		\end{aligned}
		$$ where $R_2(n)$ denotes $n$ modulo $2$.
	\end{prop}
	From the above proposition, we readily see that 
	\begin{equation}\label{eq_linear_compl}
	\lim _{n \rightarrow \infty} E\left[L_n\right] = \frac{n}{2} + c_n \text{ and } \lim _{n \rightarrow \infty} \operatorname{Var}\left[L_n\right]=\frac{86}{81},
	\end{equation} where $c_n={(4+R_2(n))}/{18}.$ From Rueppel's characterization on the general stochastic behavior of binary random sequences, 
	a similar and more important question arose: for a randomly chosen and then fixed sequence $\bs$ over $\F_q$, what is the behavior of $L_n(\bs)$?
	To settle this question, Niederreiter \cite{Niederreiter1988} developed a probabilistic theory for linear complexity and linear complexity profiles of a given sequence $\bs$ over any finite field $\F_q$.
	More specifically, he applied techniques from probability theory to dynamic systems and continued fractions, and deduced the probabilistic limit theorems for linear complexity by 
	exploiting the connection between continued fractions and linear complexity of a sequence $\bs$.  Below we first recall the connection discussed in \cite{Niederreiter1988}.
	
	Given a sequence $\bs=s_0s_1s_2\dots$, the fractional function 
	$G(x) = s_0x^{-1} + s_1x^{-2} + s_2x^{-3} + \cdots \in \F_q[x^{-1}]$
	has
	the following continued fraction expression
	\[
	G = 1/(A_1+1/(A_2 + \cdots)),
	\] where  the partial quotients $A_1,A_2,\dots$ are polynomials over $\F_q$ with positive degrees. 
	The $n$-th linear complexity of $\bs$ can be expressed as 
	$$L_n(\bs) = \sum_{i=1}^jA_i(\bs)$$ for the integer $j$ satisfying 
	\[
	\sum_{i=1}^{j-1}A_i(\bs) + \sum_{i=1}^{j}A_i(\bs) \leq n \leq  \sum_{i=1}^{j}A_i(\bs) + \sum_{i=1}^{j+1}A_i(\bs).
	\] By the study of the above continued fraction expansion, Niederreiter obtained the following result on $L_n(\bs)$.
	
	\begin{thm} Suppose $f$ is a nonnegative nondecreasing function on the positive integers with $\sum_{n=1}^{\infty}q^{-f(n)}<\infty$.
		Then we have
		\[
		\left|L_n(\bs) - \frac{n}{2}\right| \leq \frac{1}{2} f(n) \text{ for all sufficiently large } n.
		\]
	\end{thm}
	By taking $f(n) = (1+\epsilon)\log{n}/\log(q)$ for arbitrary $\epsilon >0$, a more explicit law of the 
	logarithm for the $n$-th linear complexity of a random sequence $\bs$ can be obtained as follows.
	\begin{equation}\label{eq_log_lc}
	\begin{aligned}
	& \varlimsup_{n \rightarrow \infty} \frac{L_n(\bs)-\frac{n}{2}}{\log n}=\frac{1}{2 \log q} \text {, } \\
	& \varliminf_{n \rightarrow \infty}\frac{L_n(\bs)-\frac{n}{2}}{\log n}=-\frac{1}{2 \log q}. \\
	\end{aligned}
	\end{equation}
	Consequently, we have
	\begin{equation}\label{eq_log_lc1}
	L_n(\bs) = \frac{n}{2} + O(\log(n)).
	\end{equation}
	As the equalities in (\ref{eq_log_lc}) hold
	for infinitely many $n$,
	there is a strong interest in sequences $\bs$ whose $n$-th linear complexity has small derivations from $n/2$.
	
	\begin{defn} Let $d$ be a positive integer. A sequence $\bs=s_0s_1s_2\dots$ is said to be $d$-perfect if
		\[
		\left|L_n(\bs) - \frac{n}{2}\right| \leq \frac{d}{2} \text{ for any } n\geq 1.
		\]A $1$-perfect sequence is also called perfect. A sequence is called almost perfect if
		it is d-perfect for some integer $d>1$.
	\end{defn}
	Constructing $d$-perfect sequences is of significant interest. Sophisticated methods based 
	on algebraic curves was introduced in \cite{Xing1999},
	which yielded sequences with almost perfect linear complexity profile.
	For cryptographic applications, especially as a keystream, a sequence $\bs$ should have a  linear complexity profile close to that of a random sequence. 
	Moreover, this condition should be true for any starting point of the sequence. That is to say, for a sequence $\bs=s_0s_1s_2\dots$ and any $r\geq 0$, 
	the $r$-shifted sequence $s_{r}s_{r+1}s_{r+2}\dots$ should have an acceptable linear complexity profile close to random sequences as well. 
	Niederreiter and Vielhaber \cite{Niederreiter1999a} attacked this problem with the help of continued fractions.
	An important fact is that 
	the jumps in the linear complexity proﬁle of $\bs$ are exactly the degrees of  the partial quotients $A_1,A_2,\dots$.  
	With such a connection, they proposed an algorithm to determine the linear complexity proﬁle of shifted sequences $s_{r}s_{r+1}s_{r+2}\dots$
	by investigating the corresponding continued fractions \cite{Niederreiter1999}. To be more concrete, they designed a method to calculate the continued fraction expansions
	of $G_n(x), \dots, G_1(x)$, where $G_r(x) = \sum_{j=r}^{n-1}s_{j}x^{-(j+1)}$ for $r\geq 0$.
	Later in his survey paper \cite{Niederreiter2003}, Niederreiter proposed the following open problem on $d$-perfect sequences.
	\begin{prob}
		Construct  a sequence $\bs=s_0s_1s_2\dots$ over $\F_q$ such that the $r$-shifted sequences 
		$s_{r}s_{r+1}s_{r+2}\dots$ for all $r\geq 0$ are $d$-perfect for some integer $d\geq 1$.
	\end{prob}
	
	{ The preceding discussion is mainly concerned with generic infinite-length sequences.} Note that sequences from an $m$-stage LFSR over
	$\F_q$ are periodic and their maximum period is $q^m-1$, which is achieved when the characteristic polynomial $f(x)$ is 
	a primitive polynomial in $\F_q[x]$. LFSR sequences of order $m$ with period $q^m-1$ are thus known as 
	maximum-length sequences, or m-sequences for short. For a periodic sequence $\bs=\bs_n^{\infty}$, 
	the values in its linear complexity profile will remain unchanged at a certain point. This reveals
	apparently different linear complexity profiles between an infinite-length sequence of 
	certain period and a random infinite-length sequence, which basically can be considered to have 
	arbitrarily long period. 
	
	\bigskip
	
	An important tool for the analysis of the linear complexity of $n$-periodic sequences over $\F_q$ is the \textit{discrete Fourier transform}. 
	Assume $\operatorname{gcd}(n, q)=1$, which means that there exists an $n$-th primitive root $\beta$ of 
	unity in some finite extension of $\F_q$. Then the discrete Fourier transform of a time-domain $n$-tuple $\left(a_0, \ldots, a_{n-1}\right) \in \mathbb{F}_q^n$ is the (frequency-domain) $n$-tuple $\left(b_0, \ldots, b_{n-1}\right)$ with
	$
	b_j=\sum_{i=0}^{n-1} a_i \beta^{i j}  \text { for } j=0, \ldots, n-1,
	$ i.e.,  { 
		\begin{equation}\label{Eq_DFT}
		\text{DFT}((a_0,\dots, a_{n-1}))= (a_0,\dots, a_{n-1})
		\begin{pmatrix}
		1 & \beta & \dots &  \beta^{n-1} \\ 
		1 & \beta^2 & \dots &  \beta^{2(n-1)} \\
		\vdots & \vdots & \ddots & \vdots \\
		1 & \beta^{n-1} & \dots & \beta^{{n-1}(n-1)}
		\end{pmatrix}.
		\end{equation}
	}
	In this case, the linear complexity of an $n$-periodic sequence can be determined via the discrete Fourier transform \cite{Jungnickel1993}.
	\begin{prop} Let $\bs=s_0s_1s_2\dots$ be an $n$-periodic sequence over $\F_q$ with $\gcd(n, q)=1$. 
		Let $\beta$ be an $n$-th primitive root of unity in an extension of $\F_q$ and $\left(b_0, \ldots, b_{n-1}\right)$ be the discrete Fourier transform of $(s_0,s_1,\dots s_{n-1})$ as in Eq. \eqref{Eq_DFT}. Then 
		$$
		L_n(\bs) = \#\{ j: b_j \neq 0, 0\leq j < n\} = n - \#\left\{j: \sum_{i=0}^{n-1}s_i \beta^{i j} = 0\right\}.
		$$
		
	\end{prop}
	Massey and Serconek in \cite{Massey1996} further extended the above relation to the general case $\gcd(q, n)>1$ with the generalized discrete Fourier transform. 
	The generalized discrete Fourier transform of $(s_0,s_1,\dots, s_{n-1})$, where 
	$n=p^vw$, $\gcd(w,p)=1$ and $p$ is the characteristic of $\F_q$, is defined as the following $p^v\times n$
	matrix
	$$
	\operatorname{GDFT}\left((s_0,s_1,\dots, s_{n-1})\right)=\left(\begin{array}{cccc}
	s(1) & s(\beta) & \cdots & s\left(\beta^{n-1}\right) \\
	s^{[1]}(1) & s^{[1]}(\beta) & \cdots & s^{[1]}\left(\beta^{n-1}\right) \\
	\vdots & & & \\
	s^{\left[p^v-1\right]}(1) & s^{\left[p^v-1\right]}(\beta) & \cdots & s^{\left[p^v-1\right]}\left(\beta^{n-1}\right)
	\end{array}\right)
	$$ where $\beta$ is any $n$-th primitive root of unity over $\F_q$ and 
	$$s^{[t]}(x) = \sum_{i=0}^{n-1}\binom{i}{t}s_ix^{i-t}$$ is the $t$-th Hasse derivative of 
	the polynomial $s_0+s_1x+\dots +s_{n-1}x^{n-1}$. It's clear that when $\gcd(n, q)=1$, 
	the GDFT of $(s_0,s_1,\dots, s_{n-1})$ reduced to the discrete Fourier transform of $(s_0,s_1,\dots, s_{n-1})$.
	As pointed out in \cite{Massey1994, Massey1996}, 
	the linear complexity of $\bs=\bs_n^{\infty}$ can be calculated in terms of the Günther weight of 
	$\operatorname{GDFT}\left((s_0,s_1,\dots, s_{n-1})\right)$, which was referred to as the Günther–Blahut theorem, which
	was used by Blahut implicitly in \cite{Blahut1983}.
	\begin{prop}Let $\bs=\bs_n^{\infty}$  be an $n$-periodic sequence over $\F_q$ with characteristic $p$, where $n=p^vw$ with $\gcd(w,p)=1$.
		Then the linear complexity of $\bs$ is equal to the Günther weight of the GDFT of the $n$-tuple $(s_0,s_1,\dots, s_{n-1})$, where
		the Günther weight of a matrix $M$ is  the number of its entries that are nonzero.
	\end{prop}

	Algebraically, the linear complexity of $\bs_n^{\infty}$ can be obtained via the feedback polynomial of the LFSR generating $\bs$ \cite{Selmer1966}.
	Consider the generating function 
	$$
	g(x) = \sum_{i=0}^{\infty}s_ix^i = \frac{s_0+s_1x+\dots + s_{n-1}x^{n-1}}{1-x^n}.
	$$
	Assume the characteristic polynomial of $\bs$ is given by $f(x)=x^{m}+a_{m-1}x^{m-1} + \dots + a_1x+a_0$ with $a_0=1$.
	Then 
	$$
	\begin{aligned}
	f(x)g(x) & = s_0 + (s_1+a_{1}s_0)x+\dots + (s_{m-1}+a_{1}s_{m-2} + \dots + a_{m-1}s_0)x^{m-1}
	\\ & \hspace{9.5mm} + \sum_{i\geq 0}(s_{i+m}+a_{1}s_{i+m-1} + \dots + a_ms_{i})x^{i+m} 
	\\ & = s_0 + (s_1+a_{m-1}s_0)x+\dots + (s_{m-1}+a_{m-1}s_{n-2} + \dots + a_1s_0)x^{m-1}.
	\end{aligned}
	$$ Letting $\varphi(x) = -f(x)g(x)$, we have 
	\[
	g(x)= -\frac{\varphi(x)}{f(x)} = - \frac{\varphi_0(x)}{f_0(x)},
	\]
	where $\gcd(\varphi_0, f_0)=1$ and $f_0$ is the minimal polynomial of $\bs$. This relation implies
	the following equality \cite{Selmer1966}:
	$$
	f_0(x)s_n(x) = (x^n-1)\varphi_0(x).
	$$ where $s_n(x)=s_0+s_1x+\dots + s_{n-1}x^{n-1}$.
	Since $\gcd(\varphi_0, f_0)=1$, it follows that $f_0(x) | (x^n-1)$ and $\varphi_0(x) | s_n(x)$. Therefore, we have the following result.
	\begin{prop}\label{prop_lc_algebraic} Let $\bs=\bs_n^{\infty}$ be an $n$-periodic sequence over $\F_q$ and $s(x)$ be the associated function of $\bs$ given by $s_n(x)=s_0+s_1x+\dots + s_{n-1}x^{n-1}$.
		Then the minimal polynomial of $\bs$ is given by 
		$$
		f_0(x) = \frac{x^n-1}{\gcd(x^n-1, s_n(x))},
		$$ indicating that 
		$$
		L(\bs) = n - \deg(\gcd(x^n-1, s_n(x))).
		$$
		
	\end{prop}
	With the above neat expression of $L(\bs)$ for periodic sequences $\bs_n^{\infty}$, 
	several families of sequences with nice algebraic structure were investigated, such as Legendre sequences \cite{Ding1998}, discrete logarithm function \cite{Meidl2001}, Lempel-Cohn-Eastman sequences \cite{Helleseth2003}. 
	For more information about linear complexity of sequences with special algebraic structures, readers are referred to Shparlinski's book \cite{Shparlinski2003}.

	The above discussion is concerned with the explicit calculation of an $n$-periodic sequence. The statistical behavior of a random $n$-periodic sequence over $\F_q$
	is of fundamental interest, particularly from the cryptographic perspective. Rueppel in \cite{Rueppel1986a} considered this problem: let 
	$\bs=\bs_n^{\infty}$ run through all $n$-periodic sequences over $\F_q$, what is the expected linear complexity 
	\[
	E_{q,n} = \frac{1}{q^n} \sum\limits_{\bs}L(\bs).
	\] For the case $q=2$, Rueppel showed that if $n$	is a power of $2$, then 
	$
	E_{2,n} = n-1+2^{-n}; 
	$ and if $n=2^m-1$ with a prime $m$, then $E_{2,n} \gtrapprox e^{-\frac{1}{n}}\left(n-\frac{1}{2}\right)$. Based on observations on numerical results he 
	conjectured that $E_{2,n}$ is always close to $n$. There was little progress on this conjecture until the work by Meidl and Niederreiter \cite{Meidl2002},
	in which they studied $E_{q,n}$ for an arbitrary prime power $q$ with the help of the above Günther–Blahut theorem and the analysis of cyclotomic cosets. 
	For an integer $0 \leq j < w$, the cyclotomic coset of $j$ modulo $w$ is defined as $C_j =\{jq^t\tpmod{w}: t \geq 0\}$. 
	
	\begin{thm}\label{thm_lc}
		Let $n=p^v w$, where $p$ is the characteristic of $\mathbb{F}_q, v \geq 0$, and $\operatorname{gcd}(p, w)=1$.
		Let $C_1, \ldots, C_h$ be the different cyclotomic cosets modulo $w$ and $b_i=\left|C_i\right|$ for $1 \leq i \leq h$. Then
		$$
		E_{q,n}=n-\mathlarger{\mathlarger{\sum}}\limits_{i=1}^h \frac{b_i\left(1-q^{-b_i p^v}\right)}{q^{b_i}-1}.
		$$
		
	\end{thm}
	From Theorem \ref{thm_lc}, a routine inequality scaling 
	$$\mathlarger{\sum}_{i=1}^h \frac{b_i\left(1-q^{-b_i p^v}\right)}{q^{b_i}-1} < \mathlarger{\sum}_{i=1}^h \frac{b_i}{q^{b_i}-1} <\frac{1}{q-1} \mathlarger{\sum}_{i=1}^h {b_i} = \frac{w}{q-1}$$
	implies that 
	\[
	E_{q,n} \geq n - \frac{w}{q-1}.
	\] 
	In the particular case $v=0$, 
	\[
	E_{q,n} \geq n - \frac{w}{q-1} = n(1-\frac{1}{q-1}).
	\]
	For small $q$, the above lower bounds can be further improved. For instance, for $q=2$, as the only singleton coset is $C_0=\{0\}$,
	it follows that 
	\[
	E_{2,n} \geq 
	\begin{cases}
	n - \frac{w-2}{3}  \text{ for any } n, \\
	\frac{3n-1}{4} \text{ for odd } n.
	\end{cases}
	\]
	Recall that the expected linear complexity of a random binary sequence $s_0s_1\dots s_{n-1}$ is around $\frac{n}{2}$. 
	For a periodic sequence $\bs=\bs_n^{\infty}$ from a sequence $\bs_n$ with linear complexity $l$, 
	the calculation of linear complexity of $\bs_n^{\infty}$ needs to consider the sequence $s_0s_1\dots s_{n-1}s_0\dots s_{l-1}$.
	This expansion sequence, when considered as a random sequence, would have expected complexity around $(n+l)/2$. This is somewhat consistent with 
	$E_{2,n} = (3n-1)/4$ when $l=(n-1)/2$ for odd $n$.
	Further improved results for the case $\gcd(q,n)=1$ were obtained by more detailed analysis and can be found in \cite{Meidl2002}.
	
	\section{Quadratic Complexity}\label{sec_qc}
	
	Quadratic feedback functions are the easiest nonlinear case and has been 
	discussed by some researchers \cite{Chan1990a,Youssef2000,Rizomiliotis2005}. Chan and Games in \cite{Chan1990a} studied the quadratic complexity of binary DeBruijn sequences of order $m$ (in which each binary 
	string $s_0s_1\dots s_{m-1}$ appears exactly once); Youssef and Gong characterized a jump property of the quadratic complexity 
	profile of random sequences, and Rizomiliotis et al. in \cite{Rizomiliotis2005} proposed an efficient method to calculate the quadratic complexity of binary sequences in general. 
	We first recall the definition of quadratic complexity of a sequence in the following.
	
	\begin{defn}\label{defn_qc}
		Given a binary sequence $\bs=s_0s_1s_2\dots$, its quadratic complexity is the length of the shortest FSRs with quadratic feedback functions
		$$
		f(x_0,x_1,\dots, x_{m-1}) = \sum\limits_{0\leq i,j<m} a_{i,j}x_ix_j + \sum\limits_{0\leq i<m} a_ix_i + a, \quad a_{i,j},a_i,a\in \F_2
		$$
		that can generate $\bs$. The quadratic complexity profile of the sequence $\bs$ is similarly defined as $Q_0(\bs),Q_1(\bs),Q_2(\bs),\dots$, where $Q_n(\bs)$ is the quadratic complexity of the first $n$ terms of $\bs$.
	\end{defn}
	
	For the first $n$ terms $s_0s_1\dots s_{n-1}$ of $\bs$, suppose it can be generated by an $m$-stage quadratic FSR. Following the recurrence as in Eq. \eqref{eq_FSR_recurrence}, we can obtain a system of $n-m$ equations. 
	More concretely, denote
	\[
	\begin{aligned}
	F(m) &= (a, a_0, \overbrace{a_1, a_{0,1}}, \overbrace{a_2, a_{1,2}, a_{0,2}}, \dots, \overbrace{a_{m-1}, a_{m-2,m-1}, \dots, a_{0,m-1}})^T,\\
	E(n,m) &= (s_m, s_{m+1}, \dots, s_{n-1})^T, \\
	M(n,m ) &= (\textbf{1}_{n-m}^T | B_0(m,n) | \dots | B_{m-1}(n,m)),
	\end{aligned}
	\]  where, for $0\leq i < m$, 
	$$
	B_i(n, m)=\left(\begin{array}{cccc}
	s_i & s_i s_{i-1} & \cdots & s_i s_0 \\
	s_{i+1} & s_{i+1} s_i & \cdots & s_{i+1} s_1 \\
	\vdots & \vdots & & \vdots \\
	s_{i+k-1} & s_{i+k-1} s_{i+k-2} & \cdots & s_{i+k-1} s_{k-1}
	\end{array}\right).
	$$ with $k=n-m$.
	According to the ordering in $F(m)$ and $M(n,m)$, it is easy to verify that 
	\begin{equation}\label{eq:qc}
	M(n,m) F(m) = E(n,m).
	\end{equation}
	Therefore, finding a quadratic feedback function in $m$ variables that outputs the sequence $\bs_n=s_0s_1\dots s_{n-1}$ is equivalent to solving the above 
	system of linear equations in $1+\frac{m(m+1)}{2}$ variables in $F(m)$. { Notice that the ordering of linear and quadratic terms  in $M(n,m)$ has a feature that 
		the quadratic terms $s_is_j$ with $j<i$ occur in $M(n,m)$ only after the term $s_i$ occurs. This feature facilitates the calculation of $Q_n(\bs)$ term by term as in the Berlekamp-Massey algorithm.}
	
	We first give a toy example to illustrate the above linear equations. Let $n=9$ and $m=3$. Then we have 
	\[
	\begin{aligned}
	M(n,m) &= (\textbf{1}_6^T | B_0(9,3)|B_1(9,3)|B_2(9,3))
	\\&=
	\left(\begin{array}{lllllll}
	1 & s_0 & s_1 & s_0 s_1 & s_2 & s_1 s_2 & s_0 s_2 \\
	1 & s_1 & s_2 & s_1 s_2 & s_3 & s_2 s_3 & s_1 s_3 \\
	1 & s_2 & s_3 & s_2 s_3 & s_4 & s_3 s_4 & s_2 s_4 \\
	1 & s_3 & s_4 & s_3 s_4 & s_5 & s_4 s_5 & s_3 s_5 \\
	1 & s_4 & s_5 & s_4 s_5 & s_6 & s_5 s_6 & s_4 s_6 \\
	1 & s_5 & s_6 & s_5 s_6 & s_7 & s_6 s_7 & s_5 s_7 \\
	\end{array}\right),
	\end{aligned}
	\]
	and 
	$$F(3) = (a, a_0, a_1, a_{0,1}, a_2, a_{1,2}, a_{0,2})^{T}, \quad E(9,3)=(s_3,s_4,s_5,s_6,s_7,s_8)^{T}.$$ From the relation $M(9,3) F(3) = E(9,3)$, we obtain 
	6 equations with $7$ variables in $F(3)$. 
	
	It is straightforward to see that the system is solvable if and only if $$
	\operatorname{rank}(M(n, m))=\operatorname{rank}(M(n, m) \mid E(n, m)).$$
	Chan and Games in \cite{Chan1990a} had the following observation. 
	
	\begin{thm}\label{thm_qc_1}
		If an $m$-stage quadratic FSR can generate $\bs_{n-1}$ but not $\bs_n$,
		then there is no $m$-stage quadratic FSR that can generate $\bs_n$ if and only if 
		$$\operatorname{rank}(M(n, m))= \operatorname{rank}(M(n-1, m)).$$ 
	\end{thm}
	By this theorem, for each step, when $s_0s_1\dots s_{n-2}$ has quadratic complexity $m$, if $\operatorname{rank}(M(n, m)) \neq \operatorname{rank}(M(n-1, m)),$
	then $s_0s_1\dots s_{n-2}s_{n-1}$ for any $s_{n-1}$ also has the same quadratic complexity $m$. Based on such an observation, they proposed a term-by-term
	algorithm, similarly to the Berlekamp-Massey algorithm, to calculate the quadratic complexity of a sequence $\bs_n$ \cite{Chan1990}. 
	Their algorithm strongly depends on the Gaussian elimination for the computation of $\operatorname{rank}(M(n, m))$ for each new term. The special structure of the matrix $M(n,m)$ was not well taken into consideration in their algorithm, which didn't reveal the precise increment of the quadratic complexity of $\bs_n$.
	By looking into certain structure of the matrix $M(n,m)$, Youssef and Gong \cite{Youssef2000} showed that if 
	the quadratic complexity of the first $n$ terms of a sequence $\bs$,
	is larger than $n/2$, then the first $n+1$ terms of $\bs$ has the same quadratic complexity.
	Rizomiliotis, Kolokotronis and Kalouptsidis in \cite{Rizomiliotis2005} observed the following nesting structure of the matrix $M(n,m)$:
	\[
	\begin{aligned}
	M(n, m) & =\left(\begin{array}{l}
	M(n-1, m) \\
	R_{n-m-1}(n, m)
	\end{array}\right) 
	=\left(M(n-1, m-1) \mid B_{m-1}(n, m)\right)
	\\& = \left(\begin{array}{l}
	R_0(n, m) \\
	M^{\prime}(n, m)
	\end{array}\right),
	\end{aligned}
	\]
	where $R_{n-m-1}(n,m)$ is the last row in $M(n,m)$, $R_0(n, m)$ is the starting row of $M(n,m)$, and $M'(n,m)$ 
	contains all the columns of $M(n,m+1)$ with indices of the form with indices of the form $j = \frac{b(b+1)}{2}+k$, with $0\leq k\leq b< m$, including the starting all one column
	$\mathbf{1}_{n-(m+1)}^T$
	in $M(n, m+1)$. This nesting structure played an important role in the their derivation of the following results.

	\begin{thm}\label{thm_qc_2}
		Suppose $Q_{n-1}(\bs)<Q_n(\bs)$. Then $Q_n(\bs)$ is the smallest integer $m$ such that 
		\[
		\operatorname{rank}(M(n,m)) \neq \operatorname{rank}(M(n-1,m)).
		\]
	\end{thm}

	
	\begin{figure}
		\begin{center}
			\begin{tikzpicture} 
			\tikzset{
				rect/.style={rectangle, draw, text centered},
				loop/.style={ 
					draw,
					chamfered rectangle,
					chamfered rectangle xsep=2cm
				},
			}
			
			\node[rect](rect0) at (4,9.5){$n=0, m=Q_0(\bs)
				=0
				$};
			\node[rect] (rect1) at (4,8) {$n=n+1$};
			\node[rect] (rect2) at (4,6) {$m =Q_n(\bs)=Q_{n-1}(\bs)$};
			\node[loop] (rect3) at (4,4) {$\operatorname{rank}(M(n, m))\overset{?}{=}\operatorname{rank}(M(n-1, m))$};
			\node[loop] (rect4) at (4,2) {$\operatorname{rank}(M(n, m))\overset{?}{=}\operatorname{rank}(M(n, m) | E(n, m)$};
			\node[rect ] (rect5) at (4,0) {$m=Q_n(\bs)=Q_n(\bs)+1$};
			\node[loop] (rect6) at (4,-2) {$\operatorname{rank}(M(n, m))\overset{?}{=}\operatorname{rank}(M(n-1, m))$};
			
			\draw[->] (rect0) -- (rect1);
			\draw[->] (rect1) -- (rect2);
			\draw[->] (rect2) -- (rect3);
			\draw[->] (rect3) -- (rect4); \node at (4.4, 3) {Yes};
			\draw[->] (rect4) -- (rect5); \node at (4.4, 1) {No};
			\draw[->] (rect5) -- (rect6);
			
			\draw[-] (10,8) -- (10,-2); 
			\draw[->] (10,8) -- (rect1);
			\draw[-] (rect6) -- (10,-2);  \node at (8.8,-1.8) {No};
			\draw[-] (rect6) -- (-1,-2) -- (-1,0);  \node at (-0.5,-1.8) {Yes};
			\draw[->] (-1,0) -- (rect5);
			
			\draw[->] (rect3) -- (10, 4); \node at (8.8,4.2) {No};
			\draw[->] (rect4) -- (10, 2); \node at (8.8,2.2) {Yes};
			
			\end{tikzpicture}
		\end{center}
		\caption{Recursive calculation of the quadratic complexity profile of $\bs$} \label{Fig_qc}
	\end{figure}
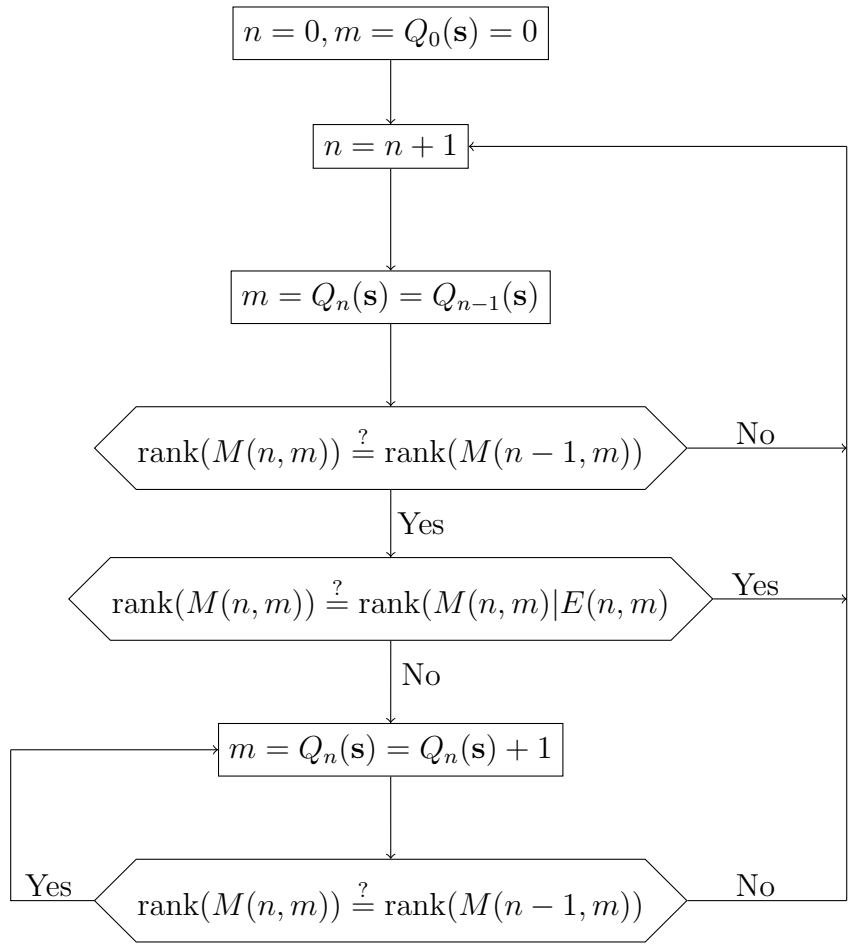

	Theorems \ref{thm_qc_1} and  \ref{thm_qc_2} laid the foundation for an algorithmic method  to calculate the quadratic complexity of a sequence.
	More specifically, for each new term,
	Theorem \ref{thm_qc_1} determines when the quadratic complexity will increase; and Theorem \ref{thm_qc_2} characterizes how large the increment is.
	Combining different cases for each new term,  Figure \ref{Fig_qc} provides a preliminary algorithm to recursively assess the quadratic complexity of a sequence $\bs$.
	It can be seen from Figure \ref{Fig_qc} that the assessment heavily depends on the calculation of ranks of involved matrices,
	which becomes slower as $n$ and $m$ grow.
	With a detailed analysis of the nesting structure of $M(n,m)$, the authors in \cite{Rizomiliotis2005} proposed a more efficient algorithm (as in  \cite[Fig. 3]{Rizomiliotis2005})
	to calculate the quadratic complexity profile of $\bs_n$. 
	In addition, when $n<\frac{m(m+3)}{2}+1$, the system $M(n,m) F(m) = E(n,m)$ is under-determined;
	and when $n\geq \frac{m(m+3)}{2}+1$, the necessary and sufficient condition that a unique quadratic feedback function generating the sequence $\bs_n$ can be given \cite{Rizomiliotis2005}.
	\begin{thm}
		The quadratic feedback function of an $m$-stage FSR that generates the sequence $\bs_n$ is unique if and only if
		$$
		\operatorname{rank}(M(n, m))=\frac{m(m+1)}{2}+1.
		$$ 
		Otherwise, there are
		$
		2^{1+\frac{m(m+1)}{2}-\operatorname{rank}(M(n, m))}
		$
		such functions.
	\end{thm}
	This theorem illustrates that a binary sequence $\bs$ with small quadratic complexity $m$ should not be used for cryptographic applications. 
	Otherwise, when $O(\frac{m(m+3)}{2})$ consective components $s_i$'s in $\bs$ are known, the quadratic feedback function could be (uniquely) determined, thereby producing the whole sequence
	and violating the requirement of unpredictability on sequences used in cryptography. 
	
	In the previous section, we saw a good theoretical understanding of the statistical behavior of linear complexity and linear complexity profile for random sequences.
	However, to the best of my knowledge, there is no published theoretical result on the statistical behavior for quadratic complexity and quadratic complexity profile of 
	random sequences $\bs_n$ and $n$-periodic sequences $\bs_{n}^{\infty}$, 
	except for the following two conjectures by Youssef and Gong \cite{Youssef2000} strongly indicated by numerical results.
	\begin{conj}
		Let $N_n(q_c)$ be the number of binary sequences $\bs_n$ with quadratic complexity $q_c=Q(\bs)>n/2$. Then $N_n(q_c) = N_{n+i}(q_c+i)$ for any $i\geq 1$, indicating that $N_n(q_c)$ is a function of $n-q_c$.
	\end{conj}
	
	\begin{conj}
		For moderately large $n$, the expected value of the quadratic complexity of a random binary sequence of length $n$ is given
		by 
		\[
		E[Q_n] \approx \sqrt{2n}.
		\]
	\end{conj}
	
	\section{Maximum-order Complexity}\label{sec_nc}
	
	As an additional figure of merit for randomness testing, Jansen and Boekee in 1989 proposed the notion of maximum-order complexity, also known as nonlinear complexity later, of sequences \cite{Jansen1990}.
	We adopt the term maximum-order complexity in this survey for better distinction with the notion of quadratic complexity.
	
	\begin{defn}\label{nlc}
		The maximum-order complexity of a sequence $\bs=s_0s_1s_2\dots$, denoted by $M(\textbf{s})$, is the length of the shortest FSRs that can generate the sequence $\textbf{s}$. 
		Similarly, the maximum-order complexity profile of $\bs$ is a sequence of $M_0(\bs),M_1(\bs), M_2(\bs), \dots, $, where $M_n(\bs)$ for each $n\geq 0$ is the 
		shortest FSRs that can generate the first $n$ terms of $\bs$.
	\end{defn}
	As pointed in \cite{Jansen1990}, the significance of the maximum-order complexity is that it tells exactly how many
	terms of $\bs$ have to be observed at least, in order to be able to generate the entire sequence by means of an FSR with length $M(\bs)$.
	Therefore, it has been considered as an important measure to judge the randomness of sequences.
	Below we recall some basic properties of maximum-order complexity of sequences \cite{Jansen1990,Jansen1991}.
	
	\begin{lem}{\label{lem_nc0}} 
		Let $\mathbf{s}=s_0s_1s_2\cdots$ be a sequence over $\F_q$.

		\noindent(i) Let $l$ be the length of the longest subsequence of $\mathbf{s}$ that occurs at least
		twice with different successors. Then the sequence $\bs$ has maximum-order complexity $M(\bs)=l+1$;
		
		\noindent(ii)  The maximum-order complexity of any $n$-length sequence is at most $n-1$. Moreover, the equality holds if and only if $\mathbf{s}_n$ has the form $({\alpha,\cdots,\alpha},\beta)$ with $\alpha\neq \beta$ in $\F_q$. 
		%
	\end{lem}
	
	\begin{lem}{\label{lem_nc1}} 
		Let $\bs=\bs_n^{\infty}$ be a sequence of period $n$ over $\F_q$. 
		
		\noindent(i) The maximum-order complexity of $\bs$ satisfies $\lceil \log_q(n)\rceil \leq M(\bs)\leq n-1$;
		
		\noindent(ii) The reciprocal sequence $\bs^*=(s_{n-1}, \dots, s_0)^{\infty}$ has $M(\bs^*)=M(\bs)$.
		%
	\end{lem}
	
	From Lemmas \ref{lem_nc0} and \ref{lem_nc1} the difference between nonlinear complexities 
	of finite-length sequences and periodic sequences is apparent. One typical difference is that the maximum-order complexity of a periodic sequence remains unchanged under 
	cyclic shift operations, while that of a finite-length sequence can change dramatically. For instance, for the sequence ${0\dots 0} 1$ of length $n$, 
	while $M(0\dots 0 1^{\infty})=M(0\dots 01) = n-1$, 
	after a right cyclic shift, we have $M(10\dots 0^{\infty}) = n-1$ but $M(10\dots 0) = 1$.
	
	Recall that for the case of linear feedback function is unique for periodic sequences, and that for quadratic feedback function
	is also unique when the matrix $M(n,m)$ with $n=\frac{m(m+3)}{2}+1$ is nonsingular, the situation for maximum-order complexity is significantly different.
	
	\begin{prop} Let $\Phi_{\bs}$ denote the class of feedback functions of the FSRs that can generate a periodic sequence $\mathbf{s}=s_n^{\infty}$ with maximum-order complexity $m$
		over $\F_q$. Then the number of functions in $\Phi_{\bs}$ is equal to $|\Phi_{\bs}| = q^{q^m-n}.$
		
	\end{prop}
	By the above proposition, the class $\Phi_{\bs}$ generally contains more than one feedback function that can generate the periodic sequence, which is similar for non-periodic sequences.
	One can search for functions exhibiting certain properties, such as the function with the least number of terms. One of the methods that minimize the number of terms and their orders in the inclusive-or sum of products of variables or their complements is to use the disjunctive normal form (DNF) representation of the feedback function. For the binary case, 
	the DNF of $f$ is given by 
	$$f(x_0,x_1,\dots, x_{m-1})=\sum_{b_0,\dots,b_{m-1} \in \F_2}f(b_0,\dots,b_{m-1} )x_0^{b_0}\cdots x_{m-1}^{b_{m-1}},$$
	where $x_i^{b_i} = x_i+b_i+1= 1$ if and only if $x_i=b_i$ for $0\leq i<m$.
	It is also interesting to note that a unique feedback function occurs for DeBruijn sequences of order $m$, which have $m$-tuples over $\F_q$ exactly once in one period $q^m$.
	For binary DeBruijn sequences of order $m$, Chan and Games in \cite{Chan1990} showed their quadratic complexity are upper bounded by $2^{m}-1-\binom{m}{2}$,
	which can be reached by those DeBruijn sequences derived from m-sequences by inserting $0$ to the all-zero subsequence of length $m-1$ in m-sequences.
	
	\subsection{Computation and Statistical Behavior}
	
	In \cite{Jansen1990} Jansen associated $M(\bs)$ with the maximum depth of a direct acyclic word graph (DAWG) from $\bs$. Below we recall a toy example in \cite{Jansen1990} to 
	generate a DAWG from a binary string, which helps readers better understand relevant notions.
	
	\begin{ex}\label{ex1}
		The sequence $\bs=110100$ has the following set of all subsequences
		\[
		\begin{aligned}
		{\rm SUB}(\bs) = \{& \lambda, 1,0,11,10,01,00,110,101,010,100,  \\
		&1101,1010,0100,11010,10100,110100\},
		\end{aligned}
		\]where $\lambda$ represents the empty sequence. For the sequence $\bs$, the endpoints 
		are given as 
		\begin{center}
			\begin{tabular}{|c|ccccccc|}
				\hline 
				$\bs$ & & 1 & 1 & 0 & 1 & 0 & 0 \\ \hline 
				endpoints & 0 & 1 & 2 & 3 & 4 & 5 & 6 \\ \hline
			\end{tabular}
		\end{center}
		For the sub-sequences of $\bs$, their endpoints are given as below
		$$
		\begin{array}{lllll}
		E(\lambda)&=\{0,1,2,3,&\hspace{-3mm} 4,5,6\}  &  \\
		E(1)&=\{1,2,4\}, &\hspace{-3mm}  E(0)&=\{3,5,6\}, \\
		E(11)&=\{2\}, &\hspace{-3mm}  E(10)&=\{3,5\}, \\
		E(01)&=\{4\}, &\hspace{-3mm}  E(00)&=\{6\}, \\
		E(110)&=\{3\}, &\hspace{-3mm}  E(101)&=\{4\}, \quad E(010)=\{5\}, \quad E(100)=\{6\}, \\
		E(1101)&=\{4\}, &\hspace{-3mm}  E(1010)&=\{5\}, \quad E(0100)=\{6\}, \\
		E(11010)&=\{5\}, &\hspace{-3mm}  E(10100)&=\{6\}, \\
		E(110100)&=\{6\} . &\hspace{-3mm} 
		\end{array}
		$$
		Subsequences are considered equivalent if they have the same set of endpoints. For instance, 
		the following are three equivalent classes with endpoints $4$, $5$ and $6$, respectively:
		\[
		\begin{aligned}
		&E_4: 01, 101, 1101 \\
		&E_5: 010,1010, 11010, \\
		&E_6: 00, 100, 0100, 10100, 110100. \\
		\end{aligned}
		\] Therefore, the total 17 subsequences in  ${\rm SUB}(\bs)$ are divided into $9$ equivalence classes. It's customary to choose the shortest subsequence as the class representative. 
		In this way, we have the following representatives:
		\[
		EQ(\bs) = \{\lambda, 1,0,11,10,01,00, 110, 010\}.
		\]
		From the above discussion, we can build a DAWG as follows: each representative in $EQ(\bs)$ is denoted by a node and $\lambda$ is considered as the source node;
		there is a directed edge between one node $v_i$ to another node $v_j$ if and only if the equivalence class of $v_i$ contains a subsequence, which  extended with the edge symbol belongs to
		the equivalence class of $v_j$; the edges are divided into primary and secondary edges: an edge is primary if and only if it belongs to a primary path (the longest path  from the source to the node)
		and a depth of a node is the length of the primary path to the node, where the length of a path is the number of edges along the path. Let $BN(\bs)$ be a subset consisting of
		nodes with more than one outgoing primary edges in $EQ(\bs)$ and 
		define the maximum depth $d(\bs) = \max_{v\in BN(\bs)}d(v)$. 
		
		In this way we obtain the DAWG of the sequence $\bs=110100$ 
		as in Figure \ref{Fig_DAWG}, where primary edges are displayed in solid arrows. From the definition of the depth of a node, we have 
		\[
		\begin{aligned}
		& d(\lambda) = 0, d(0)=d(1)=1, \\
		& d(11)=d(10)=2, d(01)=4, d(00)=6, \\
		& d(110) =3, d(010)=5.
		\end{aligned}
		\] Since $BN(\bs)=\{\lambda, 1, 0, 10 \}$, we have $d(\bs) = d(10) = 2.$
		
		\begin{center}
			\begin{figure}[t!]
				\centering\includegraphics[width=0.7\textwidth]{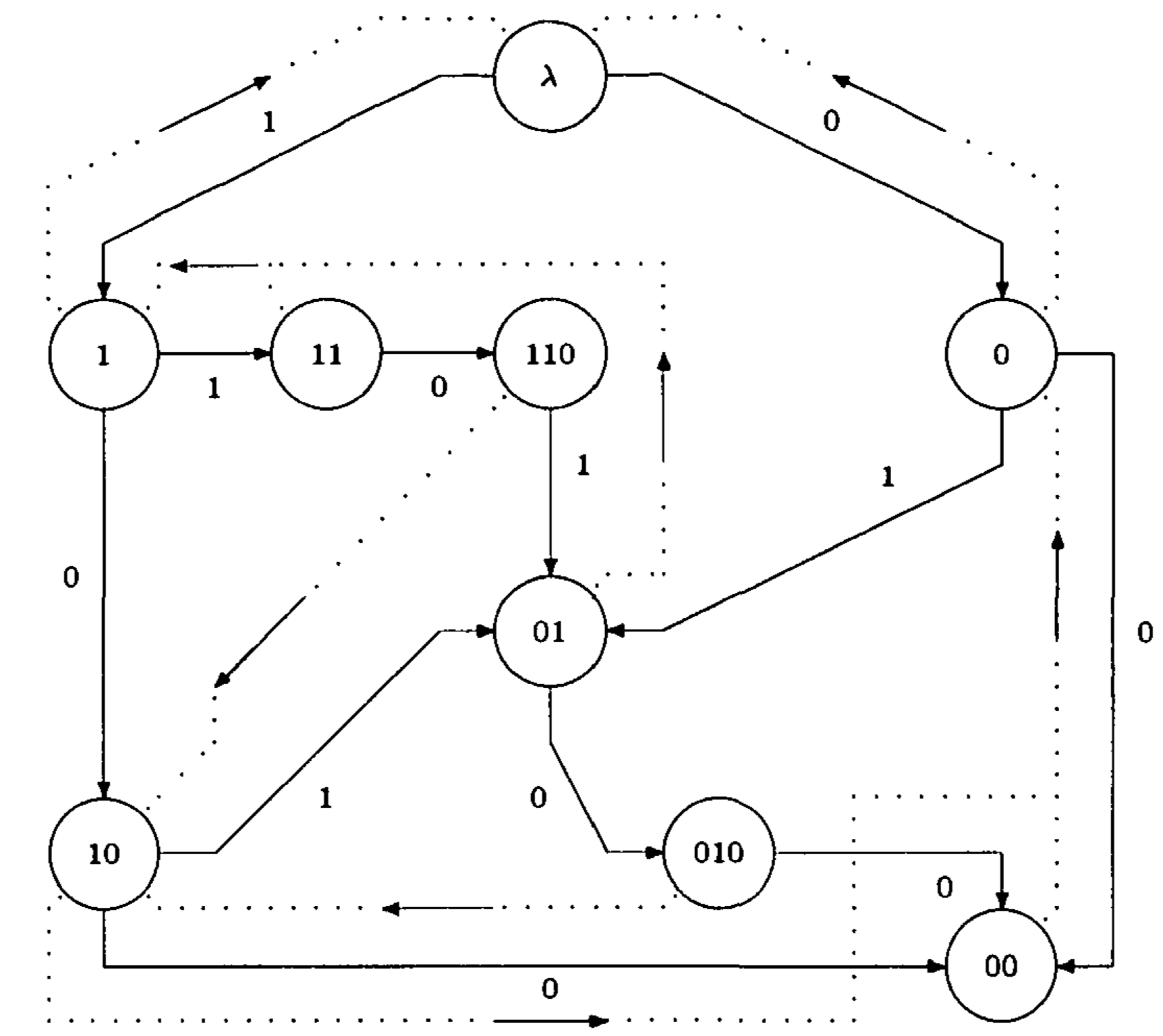}
				\caption{Direct Acyclic Word Graph of $\bs=110100$ \cite{Jansen1989} }\label{Fig_DAWG}
			\end{figure}
		\end{center}

	\end{ex}
	
	With the notions illustrated in Example \ref{ex1}, Jansen in \cite{Jansen1989} established the following connection between the maximum-order 
	complexity of a sequence $\bs$ and the depth of DAWG derived from $\bs$, and proposed to 
	calculate the maximum-order complexity of $\bs $ from its DAWG.
	
	\begin{prop} Given a sequence $\bs$, the maximum-order complexity of $\bs$ and the depth $d(\bs)$ of its DAWG satisfy
		\[
		M(\bs) = \begin{cases}
		0, & \text{if } BN(\bs) = \emptyset, \\
		d(\bs) + 1, & \text{otherwise.}
		\end{cases}
		\]
	\end{prop}
	Blumer's algorithm \cite{Blumer1983} was identified as an important tool to build an DAWG of a sequence $\bs$, thereby calculating its
	maximum-order complexity profile in linear time and memory with regard to the sequence length. The method worked particularly well for periodic sequences.
	With this algorithm, Jansen exhausted all $2^l$ binary $l$-length sequences and $l$-periodic sequences as $l$ 
	ranges from $1$ up to $24$ \cite[Tables 3.1-3.4]{Jansen1989}, which exhibited  typical statistical behaviors of maximum-order complexity profiles of random sequences: the expected maximum-order complexity of a random sequence $\bs$ of sufficiently large length $n$ over $\F_q$ is given by 
	\[
	E(M_n(\bs)) \approx 2 \log_q(n).
	\]
	Following the successful approach in \cite{Rizomiliotis2005},
	Rizomiliotis and Kalouptsidis \cite{Rizomiliotis2005_NC} 
	considered the calculation of the maximum-order complexity of a sequence $\bs$ in a similar way.
	From the recursive equations $f(s_i,s_{i+1},\dots, s_{i+m-1}) = s_{i+m}$ for $0\leq i <n-m$,
	one can obtain a similar system to linear equations 
	\[
	M(n,m) F(m) = E(n,m),
	\]where the coefficient matrix $M(n,m)$ from all binary terms $\prod_{i=0}^{m-1}x_i^{j_i}$, where $(j_0,j_1,\dots,j_{m-1})$ is any element in $\F_2^m$
	in a special ordering according to the degree of the terms.
	For instance, when $n=10$ and $m=3$, the system of linear equations is given by
	\begin{equation*}\label{key}
	M(10,3)=
	\left(\begin{array}{llllllll}
	1 & s_0 & s_1 & s_2 & s_0s_1 & s_0s_2  & s_1 s_2 &  s_0s_1 s_2  \\
	1 & s_1 & s_2 & s_3 & s_1s_2 & s_1s_3  & s_2 s_3 &  s_1s_2 s_3  \\
	1 & s_2 & s_3 & s_4 & s_2s_3 & s_2s_4  & s_3 s_4 &  s_2s_3 s_4 \\
	1 & s_3 & s_4 & s_5 & s_3s_4 & s_3s_5  & s_4 s_5 &  s_3s_4 s_5  \\
	1 & s_4 & s_5 & s_6 & s_4s_5 & s_4s_6  & s_5 s_6 &  s_4s_5 s_6 \\
	1 & s_5 & s_6 & s_7 & s_5s_6 & s_5s_7  & s_6 s_7 &  s_5s_6 s_7 \\
	1 & s_6 & s_7 & s_8 & s_6s_7 & s_6s_8  & s_7 s_8 &  s_6s_7 s_8 
	\end{array}\right)
	\end{equation*}
	with 
	\[
	\begin{aligned}
	&F(3) =(a, a_0,a_1,a_3, a_0a_1, a_0a_2, a_1a_2, a_0a_1a_2)^{T}, \\
	&E(10,3)=(s_3,s_4,s_5,s_6,s_7,s_8,s_9)^{T}.
	\end{aligned}
	\]
	In \cite{Rizomiliotis2005_NC}  the authors investigated the properties of the $m$-tuples $T_j(n,m)=(s_j,s_{j+1},\dots, s_{j+m-1})$ as $j$ runs through 
	the indices $0,1,\dots, n-m$ and proposed a term-by-term algorithm to compute the maximum-order complexity and output a feedback function for a given 
	sequence $\bs_n$. They also pointed out that the dominant multiplication complexity in their proposed algorithm is in the order of $O(n)$.
	
	\smallskip 
	
	The above approach by analyzing the structure of $M(n,m)$ is not satisfactory enough. Later in \cite{Limniotis2007} the authors 
	made more observations that further improved the performance of calculating maximum-order complexity of sequences. 
	
	\begin{prop}\label{prop_nc1}
		For a binary sequence $\bs$, suppose $M_{n-1}(\bs) = m$ and the minimal FSR of $s_0s_1\dots s_{n-2}$ does not generate $s_0s_1\dots s_{n-2} s_{n-1}$.
		Then $M_{n}(\bs)=m$ if and only if the subsequence $s_{n-m-1}s_{n-m}\dots s_{n-2}$ has not appeared within $\bs_{n-1}$.
	\end{prop}
	
	The above observation is a natural extension of the property of maximum-order complexity in Lemma \ref{lem_nc0} (i). Instead, the following observation characterized
	the jump of maximum-order complexity profile at each term.
	
	\begin{prop}\label{prop_nc}
		For a binary sequence $\bs$, suppose $M_{n-1}(\bs) = m<M_n(\bs)$. Let $i\leq n-m-1$ be the least integer such that $s_j\dots s_{j+m-1}=s_i\dots s_{i+m-1}$ for certain $0\leq j<i$.
		Then $M_{n}(\bs) = M_{n-1}(\bs) +(n-m-i) = n-i.$ Moreover, if $M_n(\bs) = m+k$ for $k\geq 1$, then the sequence $\bs_n || s_{n}\dots s_{n+k-1}$
		by arbitrary extension $s_{n}\dots s_{n+k-1}$ has the same maximum-order complexity $m+k$.	
	\end{prop}
	In addition, they observed that the maximum-order complexity profile of a sequence has a close connection to its eigenvalue profile. 
	To be more concrete, the eigenwords in a sequence $s_0s_1\dots s_{r-1}$ are those subsequences of $s_0s_1\dots s_{r-1}$ that  are not contained in any proper subsequence $s_0\dots s_{t-1}$ of $s_0s_1\dots s_{r-1}$.
	They showed the following interesting connection.
	\begin{thm}\label{thm_nc_eign}
		For a binary sequence $\bs$, suppose $M_{n-1}(\bs) = m$ and the minimal FSR of $\bs_{n-1}$ does not generate $\bs_n$. Then 
		\[
		M_n(\bs) = \max\{
		M_{n-1}(\bs), n-{\rm Eigenvalue}(\bs_{n-1}),
		\}
		\]	where ${\rm Eigenvalue}(\bs_{n-1})$ is the number of eigenwords in the sequence $\bs_{n-1}$.
	\end{thm}
	
	From the observations on the update of certain feedback function when the minimal FSR of $\bs_{n-1}$ does not generate $\bs_n$, they
	proposed the following procedure to generate a minimal feedback function and maximum-order complexity of a sequence $\bs$.
	Suppose $M_{n-1}(\bs) = m$ and the minimal feedback function of $\bs_{n-1}$ is $h_{n-1}(x_0,\dots, x_{m-1})$. Then
	\begin{itemize}
		\item if $s_{n-1}=h_{n-1}(s_{n-m-1},\dots, s_{n-2})$, then $\bs_n$ and $\bs_{n-1}$ share the minimal feedback function $h_{n-1}$ and the same maximum-order complexity $m$
		\item if $s_{n-1}\neq h_{n-1}(s_{n-m-1},\dots, s_{n-2})$ and $M_{n}(\bs)=M_{n-1}(\bs) $, then update $h_{n-1}$ as 
		$$h_{n}(x_0,\dots, x_{m-1}) = h_{n-1}(x_0,\dots, x_{m-1}) + \prod\limits_{i={n-m-1}}^{n-2}(x_i+ s_i+1)$$
		\item if $s_{n-1}\neq h_{n-1}(s_{n-m-1},\dots, s_{n-2})$ and $M_{n}(\bs)=m+k $, then 
		$$h_{n}(x_0,\dots, x_{m+k-1}) = h_{n-1}(x_0,\dots, x_{m+k-1}) + \prod\limits_{i={n-m-k-1}}^{n-2}(x_i+s_i+1)$$
	\end{itemize}
	Consequently, they proposed an efficient algorithm for maximum-order complexity, which works very similarly to the Berlekamp-Massey algorithm.
	For ease of comparison with the Berlekamp-Massey algorithm, we recall it in Algorithm \ref{alg1}.
	The computational complexity of Algorithm \ref{alg1} in 
	the worse case is $O(n^3)$, 
	while in the average case is $O(n^2\log_2(n))$ since
	the the expected maximum-order complexity $E[M_n] \approx 2\log_2(n)$. While it has the similar complexity as the DAWG method \cite{Jansen1989}, its 
	recursive nature is an important advantage since it eliminates the need to know the entire sequence in advance. This resembles the well-known Berlekamp-Massey 
	algorithm in the linear case: if a discrepancy occurs at certain step, then a well-determined corrective function is added to the current minimal feedback function of $\bs_{n-1}$
	to provide an updated minimal feedback of $\bs_n$. In addition, the update of maximum-order complexity $M_n(\bs) = \max\{
	M_{n-1}(\bs), n-{\rm Eigenvalue}(\bs_{n-1}) \}$ at each step is similar to $L_n(\bs)=\max\{L_{n-1}(\bs), n-L_{n-1}(\bs)\}$ for linear complexity of the sequence $\bs$ in the recursive procedure.
	
	\bigskip
	
	\begin{algorithm}[ht]
		\small
		\SetAlgoLined
		\KwIn{A binary sequence  $\bs=s_0s_1s_2\dots $}
		\KwOut{$ m=M_n(\bs)$\\ 
			\hspace{1.15 cm} the minimal feedback polynomial $h$ generating $\bs_n$}
		\tcp{Initialization}
		$k=0, m=0, h=s_0$ \\
		\For{$j\in[1,\dots, n-1]$}
		{   
			\tcp{check discrepancy}
			$d = s_j-h(s_{j-m}, \dots, s_{j-1})$ \\
			\eIf{$d\neq 0$}
			{
				\eIf{$m=0$}
				{
					$k=j$, $m=j$
				}
				{
					\If {$k\leq 0$}	
					{
						$t = {\rm Eigenvalue}(\bs_{n})$\\					
						\If{$t<j+1-m$}{
							$k=j+1-t-m$ \\
							$m=j+1-t$
						}
					}
					{
						$k=k-1$
					}
					
				}	$f=\prod\limits_{i=0}^{m-1}(x_i+s_{i} \oplus 1)$\\
				$h = h+f$
				
			}
			{
				$k=k-1$
			}
		}
		Return $h$
		\caption{Generation of minimal polynomial for $\bs$}\label{alg1} 
		\normalsize
	\end{algorithm}

	\subsection{Approximate statistical behavior}

	Conducting a rigorous theoretical analysis of the behavior of maximum-order complexity profiles, as done for linear complexity, appeared intractable. On the other hand,  
	maximum-order complexity had a lot of similarities as linear complexity. In order to facilitate the randomness test with maximum-order complexity, Erdman and Murphy proposed a method to 
	approximate the distribution of the maximum-order complexity \cite{Erdmann1997}. Inspired by the property of maximum-order complexity in Lemmas \ref{lem_nc0} and \ref{lem_nc1}, they investigated 
	a function that approximates the function $P(m,n)$: the probability that the ﬁrst $n$ $m$-tuples are all different in a sequence. 
	
	\begin{prop} Let $R(m,n-1)$  be the probability that the $n$-th $m$-tuple is unique given that the previous $(n-1)$ $m$-tuples were unique. Then 
		\[
		P(m,n) = \prod\limits_{i=1}^{n}R(m,i) \approx  \prod\limits_{i=1}^{n}\left(1-\frac{i+1}{2^{m+1}}\right) = p(m,n).
		\]
	\end{prop}
	Simulations on random sequences for $m$ from $4$ to $24$ \cite[Table 2]{Erdmann1997} indicated that the above approximation was accurate, particularly when $m\geq 17$.
	
	\smallskip
	
	Recall that a purely periodic sequence has maximum-order complexity $m$ if the ﬁrst $n$ $m$-tuples are unique but at least one of the ﬁrst $n$ $(m-1)$-tuples is repeated. 
	Thus, for calculating a periodic sequence $\bs=\bs_n^{\infty}$, it suffices to only look at the ﬁrst $n+m-1$ bits to see if the $n$ $m$-tuples are unique and the ﬁrst $n+m-2$ bits to see if the $n$ $(m-1)$-tuples are not unique.
	Denote by $Q(m,n)$ the probability that the first $n$ $m$-tuples are unique while the $n$ $(m-1)$-tuples are not. This probability function can be
	well approximated by $q(m,n)$, which is given by
	\[
	q(m,n) = \begin{cases}
	p(m,n) - p(m-1,n), & \text{ if } n \leq 2^{m-1}, \\
	p(m,n) & \text{ otherwise}. \\
	\end{cases}
	\] Based on this approximation (of which the accuracy was shown in \cite[Table 3]{Erdmann1997}), the expected maximum-order complexity was approximated as follows:
	
	\begin{thm}
		Let $M_n$ be the maximum-order complexity of a random periodic binary sequence $\bs_n^{\infty}$. 
		Then 
		\begin{equation*}\begin{aligned}
		E[M_n] &= \sum\limits_{m=\log_2(n)}^{n-1} m Q(m,n) = (n-1) -  \sum\limits_{m=\log_2(n)}^{n-2} P(m,n) \\
		& \approx  (n-1) -  \sum\limits_{m=\log_2(n)}^{n-2} p(m,n) \\
		&= (n-1) -  \sum\limits_{m=\log_2(n)}^{n-2}\prod\limits_{i=1}^{n}\left(1-\frac{i+1}{2^{m+1}}\right). \\
		\end{aligned}
		\end{equation*}
	\end{thm}
	
	Similarly, for random binary sequences of length $n$, the expected maximum-order complexity was approximated below.
	
	\begin{thm}
		Let $N(m,n)$ be the number of binary sequences of length $n$ with maximum-order complexity $m$. Then it can be approximated as 
		$$N(m,n) \approx 2^nq(m, n-m+1),$$ and the expected maximum-order complexity $\widehat{M}_n$ of random binary sequence of length $n$ is given by
		\[
		\begin{aligned}
		E[\widehat{M}_n]  & =\sum\limits_{m=0}^{n-1} m \frac{N(m,n)}{\sum_{i=0}^{n-1}N(i,n)} \\
		& \approx \sum\limits_{m=0}^{n-1} m \frac{2^nq(m, n-m+1)}{\sum_{i=0}^{n-1}2^nq(i, n-i+1)} \\
		& \approx 2\log_2(n)  \quad (\text{for  sufficiently large } n).
		\end{aligned}
		\]
	\end{thm}

	Motivated by pushing nonlinear complexity as a metric of sequences in statistical test as in \cite{Erdmann1997}, Petrides and Mykkeltveit \cite{Petrides2006,Petrides2008Dec} considered the classification of binary periodic sequences in terms of their nonlinear complexities. More concretely, 
	they defined a certain type of recursion of periodic binary sequences: a binary sequence $\bs_n^{\infty}$ is said to satisfy the recursion $R_{\bv}(n, i)$  if 
	$ s_{t+i} = s_t \oplus v_t
	$ for $0\leq i \leq n-1, \,0 < i \leq n-1$ and the vector $\bv = v_0v_1\dots v_{n-1}$ is the relating vector, which contains an even number of $1$'s, $v_{0}\dots v_{n-\gamma-1}=0\dots 0 1$ for certain $\gamma$ and $v_{n-1}=1.$
	They established the following connections between the above recursion and the number $N(n,m)$ of periodic binary sequences $\bs_n^{\infty}$ of nonlinear complexity $m$.
	
	\begin{thm}
		Let $V_{n,\gamma}$ be the set of $2^{\gamma-2}$ possible relating vector for given integers $n$ and  $\gamma$ and $S(R_{\bv}(n,i), m)$ be the set of cyclically inequivalent sequences of nonlinear complexity $m$ which satisfy recursion $R_{\bv}(n,i)$. Then
		one has 
		\[
		N(n,m) = \sum\limits_{\bv \in V_{n,\gamma}}\sum_{i=0}^{n-1}|S(R_{\bv}(n,i), m)|.
		\]
		In particular, when $\bv$ has the form $\bv = 0\dots 0 1\overbrace{0\dots 0}^{\gamma-2}1$ for certain $\gamma$, 
		if $g= \gcd(n, i) | \gamma$, then one has 
		\[
		|S(R(n,i,\gamma))| = 
		\begin{cases}
		2^g, & \text{ if } n\neq 2\gamma, \\
		2^{g-1}, & \text{ if } n =2\gamma.
		\end{cases}
		\]
	\end{thm}

	\subsection{Sequences with high maximum-order complexity}
	
	Different complexity measures were proposed in the literature to evaluate the randomness of sequences. 
	{ A sequence with low complexity, including the linear, quadratic and maximum-order complexity, allows for short FSRs to generate the whole period of the sequence.
		That is to say, all remaining unknown terms in this sequence can be efficiently uncovered when the feedback function and the initial state of the short FSRs are determined.
		It is clear that this kind of sequences should be avoided for any cryptographic applications. On the other hand,  the relation between high complexity of a sequence and good randomness of a sequence is not yet well understood. 
	}
	For the aforementioned complexity measures, the sequences of the form $(0,\dots, 0, 1)$ have the largest complexity, but have clearly poor randomness. 
	According to the exhaustive search results on maximum-order complexity in Tables 1-3 in \cite{Jansen1989}, while sequences $(0,\dots,0,1)$ are the only instances of $n$-length sequences that have the 
	highest possible complexity $n-1$, there are multiple $n$-periodic binary sequences having the highest maximum-order complexity $n-1$.

	Researchers have been interested in the study of sequences with high maximum-order complexity \cite{Luo2017,Liang2023},
	particularly on sequences having highest possible maximum-order complexity \cite{Rizomiliotis2006,Sun2017}.
	For the latter case, Rizomiliotis \cite{Rizomiliotis2005_NC} in 2005 proposed the following necessary and sufficient condition for an $n$-periodic binary sequence to 
	have 
	maximum-order complexity the same as its linear complexity.
	
	\begin{prop}
		For a periodic binary sequence $\bs=\bs_n^{\infty}$, let $\bs_n(x)=s_0+s_1x+\dots + s_{n-1}x^{n-1}$ be the polynomial representation of $\bs_n$,
		and let $h(x)={\gcd(x^n+1, \bs_n(x))}$, $c(x)= \bs_n(x)/h(x)$ and 
		$f(x)= {x^n+1}\big /{\gcd(x^n+1, \bs_n(x))}$.
		Then 
		\begin{equation}\label{eq_nc_lc}
		M(\bs) = \min\{n-1, L(\bs)\}
		\end{equation}
		if and only if there exists integers $0\leq p_1 <p_2 \leq n-1$ satisfying 
		\[
		f(x) | 1 + (x^{p_1}+x^{p_2}(x))c(x)
		\] when $LC(\bs)\leq n-1$.
	\end{prop}
	Based on the above  condition, different families of $n$-periodic binary sequences that have the same linear complexity and maximum-order complexity 
	were proposed in \cite{Rizomiliotis2005_NC}. Moreover, an algorithm based on Lagrange interpolation and an algorithm based on the relative shift of the component sequences
	were proposed to generate binary sequences with highest possible maximum-order complexity. These two ideas were further developed for constructing such periodic sequences with 
	period of the form $2^m-1$ \cite{Rizomiliotis2006}. Roughly 10 years later, $n$-periodic sequences with highest maximum-order complexity were revisited in \cite{Sun2017},
	which provided a complete picture of sequences with maximum-order complexity.
	The authors of \cite{Sun2017} gave the necessary and sufficient condition for $n$-periodic sequences over any alphabet to have maximum-order complexity $n-1$.
	\begin{thm}\label{thm_nc_max}
		Let $\bs=\bs_n^{\infty}$ be an $n$-periodic sequence over any field $\F$. Then
		$M(\bs) = n - 1$ if and only if there exists an integer $p$ with $1\leq p <n$ such that $s_i = s_{i+p\tpmod{n}}$ for $0\leq i<n-2$ and $s_i \neq s_{i+p\tpmod{n}}$ for $i=n-2,n-1$.
		Furthermore, 
		such a sequence can, up to shift equivalence, be represented as one of the following forms
		
		\noindent(1) $\bs_n=(\alpha^{n-1}\beta)=(\overbrace{\alpha,\dots, \alpha}^{n-1}, \beta)$ where $\alpha\neq \beta \in \F$;
		
		\noindent(2) $\mathbf{s}_n=\mathbf{s}_{r_0}=\left(\mathbf{s}_{r_1}\right)^{m_1} \mathbf{s}_{r_2}$ for certain integer $r_1 \in \mathbb{Z}_n^*$ with
		$$
		\mathbf{s}_{r_{i-1}}=\left(\mathbf{s}_{r_i}\right)^{m_i} \mathbf{s}_{r_{i+1}}, i=1,2, \cdots, k,
		$$
		and $$\mathbf{s}_{r_k}=\left((\alpha)^{r_k-1} \beta\right), \mathbf{s}_{r_{k+1}}=(\alpha),$$ where the integers $m_i, r_{i+1}$ for $i=1,2, \cdots, k$ are derived from $r_{i+1}=$ $r_{i-1}-m_i r_i$ with $r_1>r_2>\cdots>r_{k+1}=1$,
	\end{thm}
	
	With the full characterization in Theorem \ref{thm_nc_max}, the distribution of random $n$-periodic binary sequences having maximum-order complexity $n-1$
	can be derived as follows. When $q=2$, the result was derived earlier in~\cite[Th.~4]{Petrides2006}.
	  
	\begin{prop}The probability that a randomly generated sequence $\bs=\bs_n^{\infty}$ having highest $M(\bs)=n-1$ is given by
		\[
		P_{n-1} = {\rm Prob}(M(\bs)=n-1) = \frac{q(q-1)n\varphi(n)}{2\sum_{d|n}\mu(d)q^{n/d}},
		\]
		where $\varphi(\cdot)$ is the Euler's totient function and $\mu(\cdot)$ is the M\"{o}bius function given by $\mu(n)=$ $\sum_{k \in \mathbb{Z}_n^*} e^{2 \pi i k / n}$. 
		In particular, when $q=2$, the probability 
		$$P_{n-1}=\frac{\varphi(n)}{\sum_{d \mid n} \mu(d) 2^{\frac{n}{d}}}.$$
	\end{prop}
	
	Interestingly, the sequences $\bs$ characterized in Theorem \ref{thm_nc_max} exhibit a strong recursive structure, which can be derived by applying the Euclidean algorithm on $n$ and the integer $p$, a smaller period of 
	a subsequence of $\bs_n$. This strong recursive structure, on the other hand, implies that $\bs$ is far from being random. As discussed in \cite{Sun2017},
	the balancedness, stability and scalability of such sequences are not good, indicating that they should be avoided for cryptographic applications. 
	Very recently, binary sequences with periodic $n$ were further studied in \cite{Yuan2024}, where the authors further investigated the structure of $\bs_n^{\infty}$
	and proposed an algorithmic method to determine all $n$-periodic binary sequences with maximum-order complexity $\geq n/2$.
	
	\smallskip
	
	In addition, Liang et al. recently investigated the structure of $n$-length binary sequences with high maximum-order complexity 
	$\geq n/2$ and proposed an algorithm that can completely generate all those binary sequences \cite{Liang2023}. Based on the completeness, they managed to provide 
	an explicit expression of the number of $n$-length binary sequences with any maximum-order complexity between $n/2$ and $n$.
	
	\begin{prop} The number of $n$-length binary sequences with maximum-order complexity $m$ with $n/2\leq m \leq n-1$ is given by
		\[
		N(m,n) = \sum\limits_{d=1}^{n-m}(n-m-d+2)2^{n-m-d-1}N(m,m+d, d)
		\]
		where $$
		N(m,m+d, d)=2^{n-m}-\sum\limits_{t=1}^{r}(-1)^{t-1} \sum_{1\leq j_1<\dots<j_{t}}2^\frac{n-m}{p_1^{j_1}\cdots p_t^{j_t}}
		$$ when $n-m$ is factorized as  $n-m=p_1^{j_1}\cdots p_r^{j_r}$.
		
	\end{prop}

	\section{Relations between Complexity Measures}\label{sec_relation}
	
	Besides the complexity measures in the context of FSRs in previous sections,  some other measures have been discussed by researchers, such as Lempel-Ziv complexity \cite{Lempel1976}, 
	$2$-adic complexity for FSRs with carry operation \cite{Klapper1997}, $k$-error complexity, expansion complexity \cite{Merai2017}, and 
	correlation measure \cite{Isik2017,Chen2022}. This section will briefly review the relations among these measures. Below we start with the relation between the complexity measures in the previous sections, and we will be mainly focusing on extreme cases.

	\subsection{Relations between FSR-based complexities}
	From their definitions, for a sequence $\bs$, it is easily seen that 
	$$M_n(\bs)\leq Q_n(\bs) \leq L_n(\bs)\leq n.$$
	We know that the special sequence of the form $\bs_n=\alpha\dots \alpha \beta$, where $\alpha\neq \beta$, 
	has the largest linear complexity $n$, quadratic and maximum-order complexity $n-1$. 
	As for $n$-periodic sequences $\bs=\bs_n^{\infty}$ over $\F_q$, the largest linear complexity is also $n$, which can be seen from the expression
	$L(\bs) = n - \deg(\gcd(x^n-1, s_n(x)))$, as recalled in Proposition \ref{prop_lc_algebraic}. This can be used to characterize the periodic sequences with largest linear complexity.
	Assume $n=p^vw$ with $\gcd(p,w)=1$, where $p$ is the characteristic of $\F_q$. Let $C_1=\{0\}, C_2,\dots, C_h$ be the cyclotomic cosets modulo 
	$w$ relative to the power $q$. Let $\beta$ be a primitive $n$-th root of unity in an extension of $\F_q$. Then we have the following factorization of $x^n-1$ over $\F_q$:
	$$
	x^n-1 = \left( \prod_{	1\leq i\leq h} f_j(x)\right) =   \left(\prod_{1\leq i\leq h} \prod_{	j\in C_i} (x-\beta^i)\right)^{p^u},
	$$ which indicates that an $n$-periodic sequences over $\F_q$ has the largest linear complexity $n$ if and only if 
	$$
	\gcd(s_n(x), f_j) = 1 \text{ for } j = 1, 2, \dots, h.
	$$ Niederreiter in \cite{Niederreiter2003a} proved the existence of $n$-periodic sequences with the largest linear complexity $n$
	and large $k$-error linear complexity, which is defined as 
	$$L_{k}(\bs_n^{\infty}) = \min\{L(\mathbf{t}_n^{\infty}) \:|\; {\rm wt}(\bs_n-\mathbf{t}_n) \leq k  \}.
	$$ This result was obtained by finding integers $n=p^vw$ satisfying a condition, which was 
	derived from the following observation \cite{Niederreiter2003a}: let  $l=\min_{2\leq i\leq h}|C_i|$  and suppose $k$ is an integer such that 
	$\sum_{j=0}^k\binom{n}{j}(q-1)^j<q^l$, then there exists an $n$-periodic sequence $\bs$ satisfies
	\[
	L(\bs)=n \text{ and } L_{k}(\bs)  \geq n - p^v.
	\] As shown in the examples in \cite{Niederreiter2003a}, there are infinitely many primes $n$ allowing for the existence of binary $n$-periodic sequences of maximum possible linear complexity $n$.
	
	Finding $n$-periodic sequences with the largest maximum-order complexity seems more challenging, especially when $k$-error complexity is also considered. 
	As recalled in Section \ref{sec_nc}, Sun et at. \cite{Sun2017} revealed $n$-periodic sequences with the largest maximum-order complexity $n-1$, which have surprisingly strong 
	recursive structure. Besides this, little result about $k$-error maximum-order complexity and quadratic complexity has been reported. Here we propose two open problems for interested readers.
	
	\begin{prob}
		Characterize all the $n$-periodic sequences over $\F_q$ with the largest quadratic complexity $n-1$.
	\end{prob}
	
	\begin{prob}
		Characterize certain lower bound on the $k$-error maximum-order complexity of $n$-periodic sequences with maximum-order complexity $n-1$.
	\end{prob}
	
	Another extreme case for sequences is that they have the least possible complexity values. 
	For finite-length sequences, we know that the linear, quadratic and maximum-order complexity can be as small as zero, which don't appear interesting to explore.
	When $\bs$ is a periodic sequence over $\F_q$, we know that m-sequences of length $n=q^m-1$ have the least complexities
	$
	M(\bs)= Q(\bs)= L(\bs) = m
	$ and DeBruijn sequences of length $n=q^m$ have the least maximum-order complexity $M(\bs) = m.$
	In addition to the aforementioned observations, readers may wonder what else we know about their relations so far?

	Due to the limited understanding of DeBruijn sequences, there have been only partial results for the above problem. 
	It was well known \cite{Chan1982} that the linear complexity of binary DeBruijn sequences of order $m$ is between 
	$2^{m-1}+m$ and $2^m-1$, where both the lower and upper bound are attainable. However,  
	there exists no binary DeBruijn sequences with linear complexity $2^{m-1}+m+1$ \cite{Games1983}.  
	Based on evidences on existing results, Etzion in his survey made the following conjecture:
	\begin{conj}
		For any integer $2^{m-1}+m+2\leq d\leq 2^m-1$, there exists a binary DeBruijn sequence
		with minimal polynomial $(x+1)^d$.
	\end{conj}
	There is a close connection between m-sequence over $\F_q$ of order $m$ and DeBruijn sequences of order $m$.
	There have been also some works on the linear complexity of modified DeBruijn sequences, which obtained by removing one zero from
	the longest run of zeros in a DeBruijn sequence. Mayhew and Golomb \cite{Mayhew1990} showed that the minimal polynomial of binary modified DeBruijn sequences of order $m$ is 
	a product of distinct irreducible polynomials with degrees not equal to $1$ and dividing $m$. Further results on the restrictions of the degrees were developed in \cite{Kyureghyan2008,Dong2019,Wang2020}.
	
	When considering the fact for a binary sequence $\bs$ with low quadratic complexity $m$,
	any subsequence of length slightly larger than $\frac{(m+3)m}{2} + 1$ can be sufficient to recover the whole sequence $\bs$. 
	However, the topic of quadratic complexity was significantly less explored. 
	There has been little research progress on the quadratic complexity although it was already studied in 1989. 
	To the best of my knowledge, only result about the quadratic complexity of DeBruijn sequences were report by  Chan and Games \cite{Chan1990}.
	
	\begin{prop}
		Given a binary DeBruijn sequence $\bs$ of order $m\geq 3$, then 
		$$
		Q(\bs) \leq 2^m-\binom{m}{2}-1.
		$$In particular, this upper bound can be attained by the DeBruijn sequence derived from 
		an m-sequence of order by including the all zero $m$-tuple.
	\end{prop} 
	In addition, numerical results for $m=3,4,5,6$ led to a conjecture in \cite{Chan1990}: the quadratic complexity of binary DeBruijn 
	sequences of order $m\geq 3$ is lower bounded by $m+2$. Three years later this conjecture was later confirmed by Khachatrian in \cite{Khachatrian1993}.
	%

	
	\subsection{FSR-based complexities, Lempel-Ziv complexity, expansion complexity, and autocorrelation}
	This subsection will briefly review recent works on the relations among the linear, maximum-order  complexity and other significant metrics for 
	assessing pseudo-random sequences.
	
	\subsubsection{2-adic complexity}
	
	The 2-adic complexity introduced by Goresky and Klapper \cite{Klapper1995, Klapper1997} is closely related
	to the length of a shortest feedback with carry shift register (FCSR) that generates
	the sequence. The theory of 2-adic complexity of periodic sequences has been well developed.
	Given a sequence $\bs=\bs_n^{\infty}$ of period $n$, one has
	$$
	\sum\limits_{i=0}^{\infty}s_i2^i = -\frac{\sum_{i=0}^{n-1}s_i2^i}{2^n-1}=\frac{A}{q},
	$$where $0 \leq A \leq q, \operatorname{gcd}(A, q)=1$ and
	$$
	q=\frac{2^n-1}{\operatorname{gcd}\left(2^n-1, \sum_{i=0}^{n-1} s_i 2^i\right)}.
	$$ The $2$-adic complexity of $\bs$ is defined as $\Phi(\bs) = \log_2(q)$.
	In the aperiodic case, the $n$-th 2-adic complexity, denoted by $\Phi_{2,n}(\bs)$, is the binary logarithm of
	$$
	\min \left\{\max \{|f|,|q|\}: f, q \in \mathbb{Z}, q \text { odd }, q \sum_{i=0}^{n-1} s_i 2^i \equiv f \quad\tpmod{2^n}\right\}.
	$$
	There was very little work on exploring the relation between linear, quadratic and maximum-order complexities of a binary sequence 
	and its $2$-adic complexity.
	Until recently it was shown \cite{Chen2023} that the maximum-order complexity of a binary sequence $\bs$ is upper bounded 
	by its 2-adic complexity.
	
	\medskip
	
	\subsubsection{Lempel-Ziv complexity}
	
	The Lempel-Ziv complexity of a sequence is one classical complexity measure based on pattern counting, introduced by Lempel and Ziv in 1976 \cite{Lempel1976} 
	and later named after them. The parsing procedure in the calculation of Lempel-Ziv complexity laid the basis for the prominent Lempel-Ziv compression algorithms \cite{Ziv1977,Ziv1978}.
	The Lempel-Ziv complexity reflects the \textit{compressibility} of a sequence and thus has significant cryptographic interest.
	
	The Lempel-Ziv complexity measures the rate at which new patterns emerge as we move along a given sequence. For a binary sequence $\bs_n=s_0s_1\dots s_{n-1}$, we split up $\bs_n$ into adjacent blocks. By definition, the first block consists of $s_0$. If $s_0\ldots s_{m-1}$ is a union of blocks, or in other words if $s_{m-1}$ is the last bit in a block, then the next block \
	$s_{m}\ldots s_{m+k-1}$ is uniquely determined by two properties: i)
	the sequence $s_{m}\ldots s_{m+k-2}$ occurs as a subsequence in $s_1\ldots s_{m+k-3}$;
	ii) the sequence $s_{m} \ldots s_{m+k-1}$ does not occur as a subsequence  in $s_1 \ldots s_{m+k-2}$. 
	The Lempel-Ziv complexity is then the number of blocks into which $\bs_n$ is split up by this procedure.
	
	Motivated by Niederreiter's statement \cite{Niederreiter1999}, Limniotis, Kolokotronis and Kalouptsidis explored the relation between the maximum-order complexity 
	and Lempel-Ziv complexity of a sequence, which are both closely connected to the eigenvalue profile of the sequence. They pointed out that although the exhaustive history (which uniquely determines the Lempel-Ziv complexity)
	cannot fully estimate the maximum-order complexity profile and vise versa, there still exists a relationship between them in the sense that 
	the eigenvalue profile of a sequence of given Lempel-Ziv complexity, which determines the maximum-order complexity profile, is restricted rather than arbitrary. 
	They also established a lower bound on the compression ratio of sequences with a prescribed maximum-order complexity.
	
	\subsubsection{Expansion complexity}
	
	In 2012, Diem introduced the notion of expansion complexity of sequences \cite{Diem2012}, in which he showed that the sequences over ﬁnite ﬁelds with optimal linear complexity proposed by Xing and Lam via function expansion \cite{Xing1999} can be efficiently computed from a relatively short subsequence.
	
	Let $\bs=s_0s_1s_2\dots$ be a sequence over $\mathbb{F}_q$. For a positive integer $n$, the $n$ th expansion complexity $E_n(\bs)$ is 0 if $s_0=\cdots=s_{n-1}=0$ and otherwise the least total degree of a nonzero polynomial $h(x, y)=\sum_{i, j} h_{i j} x^i y^j \in \mathbb{F}_q[x, y]$ with
	$$
	h(x, G(x)) \equiv 0\left(\bmod x^n\right),
	$$
	where the total degree of $h(x, y)$ is given by
	$$
	\begin{aligned}
	\operatorname{deg}(h(x, y))= & \max _{i, j \geq 0}\left\{i+j \mid x^i y^j\right. \text { is a monomial of } 
	\left.h(x, y), h_{i j} \neq 0\right\} .
	\end{aligned}
	$$
	The expansion complexity of the sequence $\bs$ is defined as
	$$
	E(\bs)=\sup _{n \geq 1} E_n(\bs).
	$$ It can be verified that $E_n(\bs)\leq E_{n+1}(\bs)\leq E_n(\bs)+1$. In particular, if $h(x,y)$ is restricted to be irreducible, 
	then the $n$-th irreducible expansion complexity and irreducible expansion complexity  of $\bs$ can be defined accordingly.
	The relations between the linear complexity, maximum-order complexity and expansion complexity were discussed recently in \cite{Merai2017,Sun2021}.
	
	\begin{prop}
		Let $\bs$ be an ultimately periodic sequence with pre-period $u$, period $n$ and linear complexity $L(\bs)$. Then the irreducible expansion complexity
		$$
		E^*(\bs)= \begin{cases}L(\bs)+1 & \text { if } u=0 \\ L(\bs) & \text { if } u=1 \\ L(\bs)-1 & \text { if } u=2 \text { or } u>2, s_{u-1} \neq s_{u+n-1}\end{cases}
		$$
		and $E^*(\bs)<L(\bs)-1$ otherwise.
	\end{prop}
	
	\begin{prop}For any infinite sequence $\bs$ over $\mathbb{F}_q$, if $M_n(\bs)=n-k$ with $1 \leq k<\sqrt{2 n}-2$ and $n>8$, 
		then its $n$ th expansion complexity $E_n(\bs)$ is no more than $k+2$.
	\end{prop}
	
	\subsubsection{$k$-order correlation measure}
	
	Let $k$ be a positive integer. The $n$-th correlation measure of order $k$ of a binary sequence $\bs=s_0s_1s_2\dots$ is given by \cite{Brandstaetter2006}
	$$
	C_n(\bs, k)=\max _{u, \mathbf{d}}\left|\sum_{i=0}^{u-1}(-1)^{s_{i+d_1}+s_{i+d_2}+\ldots+s_{i+d_k}}\right|,
	$$
	where the maximum is taken over all $\mathbf{d}=(d_1,d_2,\dots, d_k)$ with non-negative integers $0\leq d_1<d_2<\cdots <d_k$ and $u$ such that $u+d_k<n$.
	
	The relations between the linear complexity, maximum-order complexity and correlation measure of order $k$ were explored in \cite{Brandstaetter2006,Isik2017,Chen2022}.
	
	\begin{prop}
		Given a binary sequence $\bs=s_0s_1s_2\dots$, the $n$-th linear complexity and $n$-th correlation measure of $\bs$
		satisfy 
		\[
		L_n(\bs) \geq n-\max _{1 \leq k \leq L_n(\bs)+1} C_n(\bs, k), \quad n \geq 1,
		\]
		and the $n$-th maximum-order complexity and $n$-th correlation measure of $\bs$ satisfy
		$$
		M_n(\bs) \geq N-2^{M_n(\bs)+1} \max _{1 \leq k \leq M_n(\bs)+1} C_n(\bs,k), \quad n \geq 1.
		$$
	\end{prop}
	
	The above bounds were recently improved in \cite{Chen2022}.
	
	\begin{prop}
		Given a binary sequence $\bs=s_0s_1s_2\dots$, if for some integer $t$, 
		the $n$-th linear complexity of $\bs$ satisfies $$2^{L_n(\bs)} < \binom{\lfloor \frac{n}{2}\rfloor}{t},$$
		then for some $k$ with $1<k\leq 2t$, one has 
		$$C_n(\bs, k) \geq \lceil \frac{n}{2}\rceil.$$
		In addition, if the $n$-th maximum-order complexity of $\bs$ satisfies $M_n(\bs)\leq M$, then one has
		$$
		C_n(\bs,2)\geq n-2^M+1.
		$$
	\end{prop}
	
	\section{Conclusion}\label{sec_conclusion}
	
	This survey primarily focused on the complexity measures of sequences within the domain of feedback shift registers, which are efficiently 
	computable and hold significant interest in cryptographic applications. Notable works on the computation, stochastic behavior,
	and theoretic result on these complexities were examined to a certain technical extent. Several conjectures were revisited and new open problems 
	were proposed. To offer a sounding overview of the research development on complexity measures, the survey also briefly reviewed the established relation between these complexities as 
	well as their connection with other important metrics for pseudo-random sequences, including the well-known 2-adic complexity, Lempel-Ziv complexity, expansion complexity and $k$-order correlation measure.
	The survey indicates that although the study of complexity measures of randomness traces back to the 1960s, only linear complexity and maximum-order complexity could be considered relatively well explored.
	Other complexity measures and their interrelations appear more intractable and require new tools, techniques, and theoretic studies in future researches.

\bibliography{seq_complexity.bib}

@Article{Chen2022,
  author        = {Zhixiong Chen and Ana I. G{\'{o}}mez and Domingo G{\'{o}}mez-P{\'{e}}rez and Andrew Tirkel},
  journal       = {Finite Fields and Their Applications},
  title         = {Correlation measure, linear complexity and maximum order complexity for families of binary sequences},
  year          = {2022},
  month         = {feb},
  pages         = {101977},
  volume        = {78},
  doi           = {10.1016/j.ffa.2021.101977},
  publisher     = {Elsevier {BV}},
}

@Book{Golomb2017,
  author        = {Solomon W. Golomb},
  publisher     = {World Scientific Publishing Company},
  title         = {Shift Register Sequences: Secure and Limited-access Code Generators, Efficiency Code Generators, Prescribed Property Generators, Mathematical Models (third Revised Edition)},
  year          = {2017},
  isbn          = {9789814632027},
}

@Article{Isik2017,
  author        = {Leyla I{\c{s}}{\i}k and Arne Winterhof},
  journal       = {Cryptography},
  title         = {Maximum-order complexity and correlation measures},
  year          = {2017},
  month         = {may},
  number        = {1},
  pages         = {7},
  volume        = {1},
  doi           = {10.3390/cryptography1010007},
  publisher     = {{MDPI} {AG}},
}

@InProceedings{Helleseth2003,
  author        = {T. Helleseth and Sang-Hyo Kim and Jong-Seon No},
  booktitle     = {Proceedings {IEEE} International Symposium on Information Theory},
  title         = {Linear complexity over {$GF(p)$} and trace representation of {Lempel-Cohn-Eastman} sequences},
  year          = {2003},
  publisher     = {{IEEE}},
  doi           = {10.1109/isit.2002.1023452},
}

@InCollection{Niederreiter1999,
  author        = {Harald Niederreiter},
  booktitle     = {SEquences and Their Applications (SETA)},
  publisher     = {Springer London},
  title         = {Some computable complexity measures for binary sequences},
  year          = {1999},
  pages         = {67--78},
  doi           = {10.1007/978-1-4471-0551-0_5},
}

@InProceedings{Massey1996,
  author        = {Massey, James L. and Serconek, Shirlei},
  booktitle     = {Advances in Cryptology --- CRYPTO '96},
  title         = {Linear complexity of periodic sequences: a general theory},
  year          = {1996},
  address       = {Berlin, Heidelberg},
  editor        = {Koblitz, Neal},
  pages         = {358--371},
  publisher     = {Springer Berlin Heidelberg},
  isbn          = {978-3-540-68697-2},
}

@Article{Massey1969,
  author        = {J. Massey},
  journal       = {{IEEE} Transactions on Information Theory},
  title         = {Shift-register synthesis and {BCH} decoding},
  year          = {1969},
  month         = {jan},
  number        = {1},
  pages         = {122--127},
  volume        = {15},
  doi           = {10.1109/tit.1969.1054260},
  publisher     = {Institute of Electrical and Electronics Engineers ({IEEE})},
}

@InProceedings{Rueppel1986,
  author        = {Rueppel, Rainer A.},
  booktitle     = {Advances in Cryptology --- EUROCRYPT' 85},
  title         = {Linear complexity and random sequences},
  year          = {1986},
  address       = {Berlin, Heidelberg},
  editor        = {Pichler, Franz},
  pages         = {167--188},
  publisher     = {Springer Berlin Heidelberg},
  doi           = {10.1007/3-540-39805-8_21},
  isbn          = {978-3-540-39805-9},
}

@Article{Merai2017,
  author        = {Merai, L. and Niederreiter, H. and Winterhof, A.},
  journal       = {Cryptography and Communications-Discrete-Structures Boolean Functions and Sequences},
  title         = {Expansion complexity and linear complexity of sequences over finite fields},
  year          = {2017},
  issn          = {1936-2447},
  number        = {4},
  pages         = {501-509},
  volume        = {9},
  doi           = {10.1007/s12095-016-0189-2},
  type          = {Journal Article},
}

@Article{Limniotis2007,
  author        = {Limniotis, Konstantinos and Kolokotronis, Nicholas and Kalouptsidis, Nicholas},
  journal       = {IEEE Transactions on Information Theory},
  title         = {On the nonlinear complexity and {Lempel--Ziv} complexity of finite length sequences},
  year          = {2007},
  issn          = {0018-9448},
  number        = {11},
  pages         = {4293-4302},
  volume        = {53},
  doi           = {10.1109/tit.2007.907442},
  type          = {Journal Article},
}

@InProceedings{Niederreiter2003,
  author        = {Harald Niederreiter},
  booktitle     = {Progress in Cryptology - INDOCRYPT 2003},
  title         = {Linear complexity and related complexity measures for sequences},
  year          = {2003},
  pages         = {1-17},
  publisher     = {Springer},
  series        = {Lecture Notes in Computer Science},
  chapter       = {Chapter 1},
  doi           = {10.1007/978-3-540-24582-7_1},
  isbn          = {978-3-540-20609-5 978-3-540-24582-7},
  type          = {Book Section},
}

@Article{Meidl2002,
  author        = {Meidl, W. and Niederreiter, H.},
  journal       = {IEEE Transactions on Information Theory},
  title         = {On the expected value of the linear complexity and the $k$-error linear complexity of periodic sequences},
  year          = {2002},
  issn          = {0018-9448},
  number        = {11},
  pages         = {2817-2825},
  volume        = {48},
  doi           = {10.1109/tit.2002.804050},
  type          = {Journal Article},
}

@InProceedings{Niederreiter1988,
  author        = {Niederreiter, Harald},
  booktitle     = {Advances in Cryptology --- EUROCRYPT '88},
  title         = {The probabilistic theory of linear complexity},
  year          = {1988},
  pages         = {191-209},
  publisher     = {Springer Berlin Heidelberg},
  series        = {Lecture Notes in Computer Science},
  doi           = {10.1007/3-540-45961-8_17},
  isbn          = {978-3-540-50251-7},
  type          = {Book Section},
}

@Book{Jungnickel1993,
  author    = {Jungnickel, D.},
  publisher = {B.I. Wissenschaftsverlag},
  title     = {{Finite Fields: Structure and Arithmetics}},
  year      = {1993},
  isbn      = {978-3-41116111-9},
  month     = {jan},
}

@book{Rueppel1986a,
	author = {Rainer A. Rueppel},
	doi = {10.1007/978-3-642-82865-2},
	publisher = {Springer Berlin Heidelberg},
	title = {Analysis and Design of Stream Ciphers},
	year = {1986}}

@InCollection{Massey1994,
  author     = {Massey, James L. and Serconek, Shirlei},
  booktitle  = {{Advances in Cryptology - CRYPTO'94}},
  publisher  = {Springer},
  title      = {A {Fourier} transform approach to the linear complexity of nonlinearly filtered sequences},
  year       = {1994},
  address    = {Berlin, Germany},
  isbn       = {978-3-540-48658-9},
  month      = {jul},
  pages      = {332--340},
  doi        = {10.1007/3-540-48658-5_31},
  journal    = {SpringerLink},
}

@book{Blahut1983,
	author = {Blahut, Richard E.},
	isbn = {978-0-20110102-7},
	publisher = {Addison-Wesley Publishing Company},
	title = {{Theory and Practice of Error Control Codes}},
	year = {1983}}

@Article{Chan1990,
  author     = {Chan, A. H. and Games, R. A.},
  journal    = {IEEE Transactions on Information Theory},
  title      = {On the quadratic spans of {DeBruijn} sequences},
  year       = {1990},
  issn       = {0018-9448},
  number     = {4},
  pages      = {822-829},
  volume     = {36},
  doi        = {10.1109/18.53741},
  type       = {Journal Article},
}

@book{Selmer1966,
	author = {E. S. Selmer},
	publisher = {Department of Informatics, University of Bergen},
	title = {Linear Recurrence Relations Over Finite Fields},
	year = {1966}}

@Article{Xing1999,
  author     = {Xing, Chaoping and Lam, Kwok Yan},
  journal    = {IEEE Transactions on Information Theory},
  title      = {Sequences with almost perfect linear complexity profiles and curves over finite fields},
  year       = {1999},
  month      = {may},
  number     = {4},
  pages      = {1267--1270},
  volume     = {45},
  doi        = {10.1109/18.761282},
  publisher  = {IEEE},
}

@Article{Niederreiter1999a,
  author     = {Niederreiter, Harald and Vielhaber, Michael},
  journal    = {Theoretical Computer Science},
  title      = {An algorithm for shifted continued fraction expansions in parallel linear time},
  year       = {1999},
  issn       = {0304-3975},
  month      = {sep},
  number     = {1},
  pages      = {93--104},
  volume     = {226},
  doi        = {10.1016/S0304-3975(99)00067-5},
  publisher  = {Elsevier},
}

@Article{Ding1998,
  author     = {Ding, Cunsheng and Hesseseth, T. and Shan, Weijuan},
  journal    = {IEEE Transactions on Information Theory},
  title      = {On the linear complexity of {Legendre} sequences},
  year       = {1998},
  month      = {may},
  number     = {3},
  pages      = {1276--1278},
  volume     = {44},
  doi        = {10.1109/18.669398},
  publisher  = {IEEE},
}

@Article{Meidl2001,
  author     = {Meidl, W. and Winterhof, A.},
  journal    = {IEEE Transactions on Information Theory},
  title      = {Lower bounds on the linear complexity of the discrete logarithm in finite fields},
  year       = {2001},
  month      = {nov},
  number     = {7},
  pages      = {2807--2811},
  volume     = {47},
  doi        = {10.1109/18.959261},
  publisher  = {IEEE},
}

@Book{Shparlinski2003,
  author    = {Shparlinski, Igor},
  publisher = {Birkh{\ifmmode\ddot{a}\else\"{a}\fi}user},
  title     = {{Cryptographic Applications of Analytic Number Theory}},
  year      = {2003},
  address   = {Basel, Switzerland},
  isbn      = {978-3-0348-8037-4},
  journal   = {SpringerLink},
}

@book{Berlekamp1968,
	address = {Singapore},
	author = {Berlekamp, Elwyn R.},
	doi = {10.1142/9407},
	isbn = {978-981-4635-89-9},
	month = {oct},
	publisher = {World Scientific Publishing Company},
	title = {{Algebraic Coding Theory}},
	year = {1968}}

@InProceedings{Chan1990a,
  author     = {Chan, Agnes Hui and Games, Richard A.},
  booktitle  = {Advances in Cryptology --- CRYPTO' 89 Proceedings},
  title      = {On the quadratic spans of periodic sequences},
  year       = {1990},
  pages      = {82-89},
  publisher  = {Springer},
  series     = {Lecture Notes in Computer Science},
  chapter    = {Chapter 9},
  doi        = {10.1007/0-387-34805-0_9},
  isbn       = {978-0-387-97317-3},
  type       = {Book Section},
}

@Article{Rizomiliotis2005,
  author     = {Rizomiliotis, P. and Kolokotronis, N. and Kalouptsidis, N.},
  journal    = {IEEE Transactions on Information Theory},
  title      = {{On the quadratic span of binary sequences}},
  year       = {2005},
  month      = {apr},
  number     = {5},
  pages      = {1840--1848},
  volume     = {51},
  doi        = {10.1109/TIT.2005.846428},
  publisher  = {IEEE},
}

@InProceedings{Youssef2000,
  author    = {Amr Youssef and Guang Gong},
  booktitle = {Proceeding of the 20th Biennial Symposium on Communications, Queen's university},
  title     = {On the quadratic span of binary sequences},
  year      = {2000},
  pages     = {159-163},
}

@PhdThesis{Jansen1989,
  author      = {Jansen, C. J. A.},
  school      = {Delft University of Technology},
  title       = {{Investigations on nonlinear streamcipher systems: construction and evaluation methods}},
  year        = {1989},
  type        = {{PhD Thesis}},
  institution = {Delft University of Technology},
}

@Article{Blumer1983,
  author  = {Blumer, Anselm and Blumer, Janet and Ehrenfeucht, Andrzej and Haussler, David and McConnell, Ross M.},
  journal = {Bulletin of the EATCS},
  title   = {Linear size finite automata for the set of all subwords of a word - an outline of results},
  year    = {1983},
  pages   = {12--20},
  volume  = {21},
}

@Article{Erdmann1997,
  author     = {Erdmann, Diane and Murphy, Sean},
  journal    = {Designs, Codes and Cryptography},
  title      = {An approximate distribution for the maximum order complexity},
  year       = {1997},
  issn       = {1573-7586},
  month      = {mar},
  number     = {3},
  pages      = {325--339},
  volume     = {10},
  doi        = {10.1023/A:1008295603824},
  publisher  = {Kluwer Academic Publishers},
}

@Article{Rizomiliotis2005_NC,
  author     = {Rizomiliotis, P. and Kalouptsidis, N.},
  journal    = {IEEE Transactions on Information Theory},
  title      = {Results on the nonlinear span of binary sequences},
  year       = {2005},
  month      = {apr},
  number     = {4},
  pages      = {1555--1563},
  volume     = {51},
  doi        = {10.1109/TIT.2005.844090},
  publisher  = {IEEE},
}

@Article{Rizomiliotis2006,
  author     = {Rizomiliotis, P.},
  journal    = {IEEE Transactions on Information Theory},
  title      = {Constructing periodic binary sequences with maximum nonlinear span},
  year       = {2006},
  month      = {aug},
  number     = {9},
  pages      = {4257--4261},
  volume     = {52},
  doi        = {10.1109/TIT.2006.880054},
  publisher  = {IEEE},
}

@Article{Luo2017,
  author     = {Luo, Yuan and Xing, Chaoping and You, Lin},
  journal    = {IEEE Transactions on Information Theory},
  title      = {Construction of sequences with high nonlinear complexity from function fields},
  year       = {2017},
  month      = {aug},
  number     = {12},
  pages      = {7646--7650},
  volume     = {63},
  doi        = {10.1109/TIT.2017.2736545},
  publisher  = {IEEE},
}

@Article{Liang2023,
  author     = {Liang, Sicheng and Zeng, Xiangyong and Xiao, Zibi and Sun, Zhimin},
  journal    = {IEEE Transactions on Information Theory},
  title      = {Binary sequences with length n and nonlinear complexity not less than $n/2$},
  year       = {2023},
  month      = {sep},
  number     = {12},
  pages      = {8116--8125},
  volume     = {69},
  doi        = {10.1109/TIT.2023.3316252},
  publisher  = {IEEE},
}

@Article{Sun2017,
  author     = {Sun, Zhimin and Zeng, Xiangyong and Li, Chunlei and Helleseth, Tor},
  journal    = {IEEE Transactions on Information Theory},
  title      = {Investigations on periodic sequences with maximum nonlinear complexity},
  year       = {2017},
  month      = {jun},
  number     = {10},
  pages      = {6188--6198},
  volume     = {63},
  doi        = {10.1109/TIT.2017.2714681},
  publisher  = {IEEE},
}

@Article{Lempel1976,
  author     = {Lempel, A. and Ziv, J.},
  journal    = {IEEE Transactions on Information Theory},
  title      = {On the complexity of finite sequences},
  year       = {1976},
  month      = {jan},
  number     = {1},
  pages      = {75--81},
  volume     = {22},
  doi        = {10.1109/TIT.1976.1055501},
  publisher  = {IEEE},
}

@Article{MartinLoef1966,
  author     = {Martin-L\"{o}f, Per},
  journal    = {Information and Control},
  title      = {The definition of random sequences},
  year       = {1966},
  issn       = {0019-9958},
  month      = {dec},
  number     = {6},
  pages      = {602--619},
  volume     = {9},
  doi        = {10.1016/S0019-9958(66)80018-9},
  publisher  = {Academic Press},
}

@Article{Church1940,
  author    = {Church, Alonzo},
  journal   = {Bulletin of the American Mathematical Society},
  title     = {On the concept of a random sequence},
  year      = {1940},
  issn      = {0002-9904},
  month     = {feb},
  number    = {2},
  pages     = {130--135},
  volume    = {46},
  publisher = {American Mathematical Society},
}

@Article{Mises1919,
  author     = {Mises, R. V.},
  journal    = {Mathematische Zeitschrift},
  title      = {Grundlagen der Wahrscheinlichkeitsrechnung},
  year       = {1919},
  issn       = {1432-1823},
  month      = {mar},
  number     = {1},
  pages      = {52--99},
  volume     = {5},
  doi        = {10.1007/BF01203155},
  publisher  = {Springer-Verlag},
}

@Article{Schnorr1973,
  author    = {Schnorr, C.P.},
  journal   = {Journal of Computer and System Sciences},
  title     = {Process complexity and effective random tests},
  year      = {1973},
  issn      = {0022-0000},
  month     = {aug},
  number    = {4},
  pages     = {376--388},
  volume    = {7},
  doi       = {10.1016/s0022-0000(73)80030-3},
  publisher = {Elsevier BV},
}

@Book{Knuth1997,
  author    = {Knuth, Donald E.},
  publisher = {Addison-Wesley},
  title     = {{The Art of Computer Programming, Volume 2: Seminumerical Algorithms}},
  year      = {1997},
  address   = {Boston},
  edition   = {Third},
  isbn      = {0201896842 9780201896848},
  refid     = {174763889},
}

@Article{Bassham2010,
  author  = {Bassham, Lawrence E. and Rukhin, Andrew L. and Soto, Juan and Nechvatal, James R. and Smid, Miles E. and Leigh, Stefan D. and Levenson, M. and Vangel, M. and Heckert, Nathanael A. and Banks, D. L.},
  journal = {NIST},
  title   = {A statistical test suite for random and pseudorandom number generators for cryptographic applications},
  year    = {2010},
  month   = {sep},
}

@Article{Niederreiter2003a,
  author    = {Niederreiter, H.},
  journal   = {IEEE Transactions on Information Theory},
  title     = {Periodic sequences with large $k$-error linear complexity},
  year      = {2003},
  issn      = {0018-9448},
  month     = {feb},
  number    = {2},
  pages     = {501--505},
  volume    = {49},
  doi       = {10.1109/tit.2002.807308},
  publisher = {Institute of Electrical and Electronics Engineers (IEEE)},
}

@InProceedings{Klapper1995,
  author    = {Klapper, Andrew and Goresky, Mark},
  booktitle = {Advances in Cryptology — CRYPTO’ 95},
  title     = {Cryptanalysis based on $2$-adic rational approximation},
  year      = {1995},
  pages     = {262--273},
  publisher = {Springer Berlin Heidelberg},
  doi       = {10.1007/3-540-44750-4_21},
  isbn      = {9783540447504},
  issn      = {0302-9743},
}

@Article{Klapper1997,
  author    = {Klapper, Andrew and Goresky, Mark},
  journal   = {Journal of Cryptology},
  title     = {Feedback shift registers, 2-adic span, and combiners with memory},
  year      = {1997},
  issn      = {1432-1378},
  month     = {mar},
  number    = {2},
  pages     = {111--147},
  volume    = {10},
  doi       = {10.1007/s001459900024},
  publisher = {Springer Science and Business Media LLC},
}

@Article{Mayhew1990,
  author    = {Mayhew, G.L. and Golomb, S.W.},
  journal   = {IEEE Transactions on Information Theory},
  title     = {Linear spans of modified {DeBruijn} sequences},
  year      = {1990},
  issn      = {0018-9448},
  number    = {5},
  pages     = {1166--1167},
  volume    = {36},
  doi       = {10.1109/18.57220},
  publisher = {Institute of Electrical and Electronics Engineers (IEEE)},
}

@Article{Wang2020,
  author    = {Wang, Hong-Yu and Zheng, Qun-Xiong and Wang, Zhong-Xiao and Qi, Wen-Feng},
  journal   = {Finite Fields and Their Applications},
  title     = {The minimal polynomials of modified {DeBruijn} sequences revisited},
  year      = {2020},
  issn      = {1071-5797},
  month     = {dec},
  pages     = {101735},
  volume    = {68},
  doi       = {10.1016/j.ffa.2020.101735},
  publisher = {Elsevier BV},
}

@Article{Kyureghyan2008,
  author    = {Kyureghyan, Gohar M.},
  journal   = {Discrete Applied Mathematics},
  title     = {Minimal polynomials of the modified {DeBruijn} sequences},
  year      = {2008},
  issn      = {0166-218X},
  month     = {may},
  number    = {9},
  pages     = {1549--1553},
  volume    = {156},
  doi       = {10.1016/j.dam.2006.11.019},
  publisher = {Elsevier BV},
}

@Article{Chan1982,
  author    = {Chan, Agnes Hui and Games, Richard A and Key, Edwin L},
  journal   = {Journal of Combinatorial Theory, Series A},
  title     = {On the complexities of {DeBruijn} sequences},
  year      = {1982},
  issn      = {0097-3165},
  month     = {nov},
  number    = {3},
  pages     = {233--246},
  volume    = {33},
  doi       = {10.1016/0097-3165(82)90038-3},
  publisher = {Elsevier BV},
}

@Article{Games1983,
  author    = {Games, Richard A},
  journal   = {Journal of Combinatorial Theory, Series A},
  title     = {There are no {DeBruijn} sequences of span $n$ with complexity $2^{n - 1} + n + 1$},
  year      = {1983},
  issn      = {0097-3165},
  month     = {mar},
  number    = {2},
  pages     = {248--251},
  volume    = {34},
  doi       = {10.1016/0097-3165(83)90060-2},
  publisher = {Elsevier BV},
}

@Article{Dong2019,
  author    = {Dong, Yu-Jie and Tian, Tian and Qi, Wen-Feng and Wang, Zhong-Xiao},
  journal   = {Finite Fields and Their Applications},
  title     = {New results on the minimal polynomials of modified {DeBruijn} sequences},
  year      = {2019},
  issn      = {1071-5797},
  month     = {nov},
  pages     = {101583},
  volume    = {60},
  doi       = {10.1016/j.ffa.2019.101583},
  publisher = {Elsevier BV},
}

@Article{Ziv1977,
  author    = {Ziv, J. and Lempel, A.},
  journal   = {IEEE Transactions on Information Theory},
  title     = {A universal algorithm for sequential data compression},
  year      = {1977},
  issn      = {0018-9448},
  month     = {may},
  number    = {3},
  pages     = {337--343},
  volume    = {23},
  doi       = {10.1109/tit.1977.1055714},
  publisher = {Institute of Electrical and Electronics Engineers (IEEE)},
}

@Article{Ziv1978,
  author    = {Ziv, J. and Lempel, A.},
  journal   = {IEEE Transactions on Information Theory},
  title     = {Compression of individual sequences via variable-rate coding},
  year      = {1978},
  issn      = {0018-9448},
  month     = {sep},
  number    = {5},
  pages     = {530--536},
  volume    = {24},
  doi       = {10.1109/tit.1978.1055934},
  publisher = {Institute of Electrical and Electronics Engineers (IEEE)},
}

@Article{Diem2012,
  author    = {Diem, Claus},
  journal   = {LMS Journal of Computation and Mathematics},
  title     = {On the use of expansion series for stream ciphers},
  year      = {2012},
  issn      = {1461-1570},
  month     = {sep},
  pages     = {326--340},
  volume    = {15},
  doi       = {10.1112/s146115701200109x},
  publisher = {Wiley},
}

@Article{Sun2021,
  author    = {Sun, Zhimin and Zeng, Xiangyong and Li, Chunlei and Zhang, Yi and Yi, Lin},
  journal   = {IEEE Transactions on Information Theory},
  title     = {The expansion complexity of ultimately periodic sequences over finite fields},
  year      = {2021},
  issn      = {1557-9654},
  month     = {nov},
  number    = {11},
  pages     = {7550--7560},
  volume    = {67},
  doi       = {10.1109/tit.2021.3112824},
  publisher = {Institute of Electrical and Electronics Engineers (IEEE)},
}

@Article{Brandstaetter2006,
  author    = {Brandstätter, Nina and Winterhof, Arne},
  journal   = {Periodica Mathematica Hungarica},
  title     = {Linear complexity profile of binary sequences with small correlation measure},
  year      = {2006},
  issn      = {1588-2829},
  month     = {jun},
  number    = {2},
  pages     = {1--8},
  volume    = {52},
  doi       = {10.1007/s10998-006-0008-1},
  publisher = {Springer Science and Business Media LLC},
}

@Article{Chen2023,
  author        = {Zhiru Chen and Zhixiong Chen and Jakob Obrovsky and Arne Winterhof},
  journal       = {arXiv},
  title         = {Maximum-order complexity and $2$-adic complexity},
  year          = {2023},
  archiveprefix = {arXiv},
  eprint        = {2309.12769},
  primaryclass  = {cs.IT},
}

@InProceedings{Topuzoglu,
  author    = {Topuzoğlus, Alev and Winterhof, Arne},
  booktitle = {Topics in Geometry, Coding Theory and Cryptography},
  title     = {Pseudorandom Sequences},
  year      = {2006},
  editor    = {Garcia, A. and Stichtenoth, H.},
  pages     = {135--166},
  publisher = {Springer},
  doi       = {10.1007/1-4020-5334-4_4},
  isbn      = {9781402053337},
  volumn    = {6},
}

@Article{Khachatrian1993,
  author    = {Khachatrian, Levon H.},
  journal   = {Designs, Codes and Cryptography},
  title     = {The lower bound of the quadratic spans of {DeBruijn} sequences},
  year      = {1993},
  issn      = {1573-7586},
  month     = {mar},
  number    = {1},
  pages     = {29--32},
  volume    = {3},
  doi       = {10.1007/bf01389353},
  publisher = {Springer Science and Business Media LLC},
}

@Article{Yuan2024,
  author    = {Yuan, Qin and Li, Chunlei and Zeng, Xiangyong and Helleseth, Tor and He, Debiao},
  journal   = {IEEE Transactions on Information Theory},
  title     = {Further investigations on nonlinear complexity of periodic binary sequences},
  year      = {2024},
  issn      = {1557-9654},
  pages     = {1--1},
  doi       = {10.1109/tit.2024.3394538},
  publisher = {Institute of Electrical and Electronics Engineers (IEEE)},
}

@InProceedings{Jansen1990,
  author     = {Cees Jansen and Dick Boekee},
  booktitle  = {Advances in Cryptology -- CRYPTO'89 Proceedings},
  title      = {The shortest feedback shift register that can generate a given sequence},
  year       = {1990},
  pages      = {90--99},
  publisher  = {Springer},
  series     = {Lecture Notes in Computer Science},
  doi        = {10.1007/0-387-34805-0_10},
  isbn       = {978-0-387-97317-3},
  type       = {Book Section},
}

@InProceedings{Jansen1991,
  author     = {Cees Jansen},
  booktitle  = {Advances in Cryptology --- EUROCRYPT '91},
  title      = {The maximum order complexity of sequence ensembles},
  year       = {1991},
  pages      = {153--159},
  publisher  = {Springer},
  series     = {Lecture Notes in Computer Science},
  chapter    = {Chapter 13},
  doi        = {10.1007/3-540-46416-6_13},
  isbn       = {978-3-540-54620-7},
  type       = {Book Section},
}

@Article{Meidl2002a,
  author    = {Meidl, Wilfried and Niederreiter, Harald},
  journal   = {Journal of Complexity},
  title     = {Linear complexity, $k$-error linear complexity, and the discrete {Fourier} transform},
  year      = {2002},
  issn      = {0885-064X},
  month     = {mar},
  number    = {1},
  pages     = {87--103},
  volume    = {18},
  doi       = {10.1006/jcom.2001.0621},
  publisher = {Elsevier BV},
}

@Article{Barker2015Jun,
  author  = {Barker, Elaine and Kelsey, John},
  journal = {NIST(National Institute of Standards and Technology), Special Publication},
  title   = {Recommendation For Random Number Generation Using Deterministic Random Bit Generators},
  year    = {2015},
  month   = {jun},
  doi     = {10.6028/NIST.SP.800-90Ar1},
}

@InProceedings{ristenpart2010,
  author    = {Ristenpart, Thomas and Yilek, Scott},
  booktitle = {Network and Distributed System Security (NDSS) Symposium},
  title     = {When good randomness goes bad: virtual machine reset vulnerabilities and hedging deployed cryptography.},
  year      = {2010},
}

@InProceedings{Woodage2019,
  author    = {Woodage, Joanne and Shumow, Dan},
  booktitle = {Advances in Cryptology -- EUROCRYPT 2019},
  title     = {An analysis of {NIST SP 800-90A}},
  year      = {2019},
  address   = {Cham},
  editor    = {Ishai, Yuval and Rijmen, Vincent},
  pages     = {151--180},
  publisher = {Springer International Publishing},
  doi       = {10.1007/978-3-030-17656-3_6},
  type      = {Book Section},
}

@InProceedings{Hannah2024,
  author    = {Davis, Hannah and Green, Matthew D. and Heninger, Nadia and Ryan, Keegan and Suhl, Adam},
  booktitle = {Public-Key Cryptography -- PKC 2024},
  title     = {On the possibility of a backdoor in the {Micali-Schnorr} generator},
  year      = {2024},
  address   = {Cham},
  editor    = {Tang, Qiang and Teague, Vanessa},
  pages     = {352--386},
  publisher = {Springer Nature Switzerland},
  doi       = {10.1007/978-3-031-57718-5_12},
  isbn      = {978-3-031-57718-5},
}

@Article{DualEC,
  author       = {Daniel J. Bernstein and Tanja Lange and Ruben Niederhagen},
  title        = {Dual {EC}: a standardized back door},
  year         = {2015},
  howpublished = {Cryptology ePrint Archive, Paper 2015/767},
  url          = {https://eprint.iacr.org/2015/767},
}

@Inbook{Kuznetsov2022,
	author="Kuznetsov, Alexandr Alexandrovich
	and Potii, Oleksandr Volodymyrovych
	and Poluyanenko, Nikolay Alexandrovich
	and Gorbenko, Yurii Ivanovich
	and Kryvinska, Natalia",
	title="Areas of Application for Nonlinear Shift Registers in {PRS} Generators",
	bookTitle="Stream Ciphers in Modern Real-time IT Systems: Analysis, Design and Comparative Studies",
	year="2022",
	publisher="Springer International Publishing",
	address="Cham",
	pages="295--318",
	isbn="978-3-030-79770-6",
	doi="10.1007/978-3-030-79770-6_12",
}

@InProceedings{Sarkar2006,
  author    = {Mukhopadhyay, Sourav and Sarkar, Palash},
  booktitle = {Computational Science and Its Applications - ICCSA 2006},
  title     = {Application of {LFSRs}for parallel sequence generation in cryptologic algorithms},
  year      = {2006},
  address   = {Berlin, Heidelberg},
  editor    = {Gavrilova, Marina and Gervasi, Osvaldo and Kumar, Vipin and Tan, C. J. Kenneth and Taniar, David and Lagan{\'a}, Antonio and Mun, Youngsong and Choo, Hyunseung},
  pages     = {436--445},
  publisher = {Springer Berlin Heidelberg},
  doi       = {10.1007/11751595_47},
  isbn      = {978-3-540-34076-8},
}

@Article{Gustavson1976May,
  author    = {Gustavson, F. G.},
  journal   = {IBM J. Res. Dev.},
  title     = {Analysis of the {Berlekamp-Massey} linear feedback shift-register synthesis algorithm},
  year      = {1976},
  month     = {may},
  number    = {3},
  pages     = {204--212},
  volume    = {20},
  doi       = {10.1147/rd.203.0204},
  publisher = {IBM},
}

@article{Kolmogorov1998Nov,
	author = {Kolmogorov, A. N.},
	title = {{On tables of random numbers}},
	journal = {Theoret. Comput. Sci.},
	volume = {207},
	number = {2},
	pages = {387--395},
	year = {1998},
	month = {nov},
	issn = {0304-3975},
	publisher = {Elsevier},
	doi = {10.1016/S0304-3975(98)00075-9}
}

@article{Dornstetter1987May,
	author = {Dornstetter, J.},
	title = {On the equivalence between {Berlekamp's and Euclid's algorithms }},
	journal = {IEEE Transactions on Information Theory},
	volume = {33},
	number = {3},
	pages = {428--431},
	year = {1987},
	month = {may},
	publisher = {IEEE},
	doi = {10.1109/TIT.1987.1057299}
}

@incollection{Petrides2006,
	author = {Petrides, George and Mykkeltveit, Johannes},
	title = {{On the Classification of Periodic Binary Sequences into Nonlinear Complexity Classes}},
	booktitle = {{Sequences and Their Applications - SETA 2006}},
	journal = {SpringerLink},
	pages = {209--222},
	year = {2006},
	publisher = {Springer},
	address = {Berlin, Germany},
	doi = {10.1007/11863854_18}
}

@article{Petrides2008Dec,
	author = {Petrides, George and Mykkeltveit, Johannes},
	title = {{Composition of recursions and nonlinear complexity of periodic binary sequences}},
	journal = {Designs, Codes and Cryptography},
	volume = {49},
	number = {1},
	pages = {251--264},
	year = {2008},
	month = {dec},
	publisher = {Springer US},
	doi = {10.1007/s10623-008-9178-6}
}
\bibliographystyle{IEEEtran}

\end{document}